\documentclass[twocolumn]{aastex631}
\hypersetup{linkcolor=red,citecolor=blue,filecolor=cyan,urlcolor=magenta}

\usepackage{acronym}  
\usepackage{amssymb} 
\usepackage{amsbsy}
\usepackage{amsmath}
\usepackage{array}
\usepackage{longtable}
\usepackage{multirow}
\usepackage{units}

\acrodef{PMPS}[PMPS]{Parkes Multibeam Pulsar Survey}
\acrodef{SMPS}[SMPS]{Swinburne Intermediate-latitude Pulsar Survey}
\acrodef{HTRU}[HTRU]{High Time Resolution Universe}
\acrodef{ISM}[ISM]{interstellar medium}
\acrodef{CNN}[CNN]{convolutional neural network}
\acrodef{MDN}[MDN]{mixture density network}
\acrodef{GMM}[GMM]{Gaussian-mixture model}
\acrodef{ReLU}[ReLU]{rectified linear unit}
\acrodef{SBI}[SBI]{simulation-based inference}
\acrodef{MCMC}[MCMC]{Markov chain Monte Carlo}
\acrodef{DM}[DM]{dispersion measure}
\acrodef{SNR}[SNR]{signal-to-noise ratio}
\acrodef{NPE}[NPE]{neural posterior estimation}
\acrodef{TSNPE}[TSNPE]{truncated sequential neural posterior estimation}
\acrodef{NLE}[NLE]{neural likelihood estimation}
\acrodef{NRE}[NRE]{neural ratio estimation}
\acrodef{CI}[CI]{credible interval}
\acrodef{HDR}[HDR]{highest-density region}
\acrodef{KDE}[KDE]{kernel-density estimation}
\acrodef{ISM}[ISM]{interstellar medium}
\acrodef{LPT}[LPT]{long period transient}

\def\msun{M$_{\odot}$\,}

\newcommand{\bt}{\boldsymbol{\theta}}

\newcommand{\bx}{\boldsymbol{x}}
\newcommand{\pr}{\mathcal{P}}

\submitjournal{ApJ} 

\shorttitle{Connecting radio pulsars, magnetars, and XDINSs in a unified evolutionary framework}
\shortauthors{Ronchi et al.}

\begin{document}

\title{Connecting radio pulsars, magnetars, and XDINSs in a unified evolutionary framework using simulation-based inference}



\correspondingauthor{Michele Ronchi}
\email{ronchi@astron.nl}

\author[0000-0003-2781-9107]{Michele Ronchi}
\affiliation{ASTRON, the Netherlands Institute for Radio Astronomy, Oude Hoogeveensedijk 4, 7991 PD Dwingeloo, The Netherlands}
\affiliation{Institute of Space Sciences (CSIC-ICE), Campus UAB, Carrer de Can Magrans s/n, 08193, Barcelona, Spain}
\affiliation{Institut d'Estudis Espacials de Catalunya (IEEC), Carrer Gran Capit\`a 2--4, 08034 Barcelona, Spain}

\author[0000-0002-8118-255X]{Celsa Pardo-Araujo}
\affiliation{Institute of Space Sciences (CSIC-ICE), Campus UAB, Carrer de Can Magrans s/n, 08193, Barcelona, Spain}
\affiliation{Institut d'Estudis Espacials de Catalunya (IEEC), Carrer Gran Capit\`a 2--4, 08034 Barcelona, Spain}

\author[0000-0002-6558-1681]{Vanessa Graber}
\affiliation{Department of Physics, Royal Holloway, University of London, Egham, TW20 0EX, UK}

\author[0000-0003-2177-6388]{Nanda Rea} 
\affiliation{Institute of Space Sciences (CSIC-ICE), Campus UAB, Carrer de Can Magrans s/n, 08193, Barcelona, Spain}
\affiliation{Institut d'Estudis Espacials de Catalunya (IEEC), Carrer Gran Capit\`a 2--4, 08034 Barcelona, Spain}

\author[0000-0003-0554-7286]{Clara Dehman} 
\affiliation{Departament de Física, Universitat d'Alacant, Ap. Correus 99, E-03080, Alacant, Spain}

\author[0000-0001-5438-0908]{Davide De Grandis} 
\affiliation{Institute of Space Sciences (CSIC-ICE), Campus UAB, Carrer de Can Magrans s/n, 08193, Barcelona, Spain}
\affiliation{Institut d'Estudis Espacials de Catalunya (IEEC), Carrer Gran Capit\`a 2--4, 08034 Barcelona, Spain}

\author[0000-0001-5674-4664]{Alessio Marino} 
\affiliation{Institute of Space Sciences (CSIC-ICE), Campus UAB, Carrer de Can Magrans s/n, 08193, Barcelona, Spain}
\affiliation{Institut d'Estudis Espacials de Catalunya (IEEC), Carrer Gran Capit\`a 2--4, 08034 Barcelona, Spain}
\affiliation{Departamento de Física, Universidad de Santiago de Chile (USACH), Av. Víctor Jara 3493, Estación Central, Chile}
\affiliation{Center for Interdisciplinary Research in Astrophysics and Space Sciences (CIRAS), Universidad de Santiago de Chile}
\affiliation{INAF - IASF Palermo, via Ugo La Malfa 153, I-90146 - Palermo, Italy}

\author[0000-0001-7611-1581]{Francesco Coti Zelati} 
\affiliation{Institute of Space Sciences (CSIC-ICE), Campus UAB, Carrer de Can Magrans s/n, 08193, Barcelona, Spain}
\affiliation{Institut d'Estudis Espacials de Catalunya (IEEC), Carrer Gran Capit\`a 2--4, 08034 Barcelona, Spain}

\author[0000-0001-8503-6958]{Joeri van Leeuwen}
\affiliation{ASTRON, the Netherlands Institute for Radio Astronomy, Oude
Hoogeveensedijk 4, 7991 PD Dwingeloo, The Netherlands}
\affiliation{Leiden Observatory, Leiden University, Einsteinweg 55, 2333 CC Leiden, The Netherlands}
\affiliation{Leiden Institute for Advanced Computer Science (LIACS), Leiden University, Einsteinweg 55, 2333 CC Leiden, The Netherlands}

\author[0000-0003-1018-8126]{José A. Pons}
\affiliation{Departament de Física, Universitat d'Alacant, Ap. Correus 99, E-03080, Alacant, Spain}


\begin{abstract}

Understanding the Galactic population of isolated neutron stars within a unified framework provides key insights into their birth properties, evolutionary pathways, and the connections between different neutron star classes. In this work, we aim to reproduce both the observed radio and quiescent X-ray emission from isolated neutron stars, including radio pulsars,  magnetars and X-ray dim isolated neutron stars (XDINSs). We develop a comprehensive population synthesis framework that models neutron star birth properties, their dynamical, rotational, and magneto-thermal evolution, as well as radio and X-ray emission, and the selection effects of corresponding surveys. To simulate realistic X-ray spectra, we account for magnetospheric resonant cyclotron scattering and interstellar absorption. Additionally, we model the observational bias introduced by magnetar outbursts, by linking the outburst rate to magnetic stresses in the stellar crust. We then employ a simulation-based inference method, namely truncated sequential neural posterior estimation, to reconstruct the birth properties, such as the initial magnetic field distribution. We find that the Galactic neutron star population can be described by a two-component log-normal distribution of birth magnetic fields with peaks at $5 \times 10^{12}$\,G and $10^{14}$\,G, respectively. We further find that the Galactic neutron star birth rate is around 3–4 per century. Our results help establish the contribution of neutron stars as central engines of powerful astrophysical transient phenomena, such as fast radio bursts, super-luminous supernovae and gamma ray bursts.

\end{abstract}




\section{Introduction}
\label{sec:intro}

Neutron stars are the compact remnants left over after the core-collapse supernovae of massive stars. By combining extraordinary properties such as extreme gravity, ultra-strong magnetic fields, and rapid rotation, neutron stars represent unique laboratories to study the behavior of matter and radiation in extreme environments. Neutron stars have been observed to emit across the entire electromagnetic spectrum from radio to gamma-rays and showcase diverse emission behavior, from regular periodic pulses to more sporadic highly energetic bursts. Due to this variety of properties and behavior, neutron stars have been classified into different classes \citep[see, e.g.,][]{Kaspi2010, Harding2013, Borghese2023, Popov2023}. The bulk of known neutron stars have been discovered as isolated radio pulsars with estimated dipolar magnetic fields in the range $\sim\unit[10^{12}-10^{13}]{G}$. Decades of radio pulsar surveys have produced a statistically rich sample, now comprising over four thousand objects \citep[see ATNF Pulsar Catalogue\footnote{\url{https://www.atnf.csiro.au/research/pulsar/psrcat/}},][]{Manchester2005}.
Among other classes of isolated neutron stars there are magnetars and X-ray Dim Isolated Neutron Stars (XDINSs). Magnetars are the neutron stars with the strongest magnetic fields exceeding $\unit[10^{14}]{G}$. Both their persistent and bursting emission is believed to be powered by the dissipation and instability of these strong magnetic fields \citep[see][for a review]{Turolla2015, Kaspi2017, Esposito2021, Rea2025}. XDINSs, also known as the ``Magnificent Seven'', are a class of isolated neutron stars with faint X-ray emission, showing an almost perfect thermal spectrum with broad absorption features \citep[see][for a review]{Turolla2009}. Owing to their low luminosities, XDINSs are detectable only within a few hundred parsecs, which is why the currently known sources have estimated distances lower than 500 pc. If considered unrelated to one another, the existence of several neutron star classes challenges the limits of the core-collapse supernova rate in our Galaxy \citep{Keane2008, Rozwadowska2021, Pardo-Araujo2026}. 
Furthermore, in recent years the borders between the different classes of isolated neutron stars have started to blur. For example, magnetar-like burst activity has been discovered in some rotation-powered pulsars (RPPs) \citep[][]{Gavriil2008, Archibald2016}, whereas pulsed radio emission was detected from several magnetars in outburst \citep[e.g.][]{Camilo2006}. Furthermore, magnetar-like and faint radio emission have been detected from yet another class, so-called central compact objects (CCOs) \citep{Rea2016, Dai2016,ZhangL2025}. These are young neutron stars found at the geometrical center of supernova remnants, which have previously been detected only as stable X-ray emitters. These observational hybrid properties indicate that the different classes of isolated neutron stars are likely connected through evolutionary pathways, and therefore should be studied as a single population rather than as distinct, unrelated categories. A unified approach helps us to better understand the neutron star birth properties, the relationship with their massive progenitors and supernova explosions that lead to their formation.

In general, the neutron star distribution in the spin period derivative, $\dot{P}$, \textit{vs} spin period, $P$, plane encodes information about the initial spin and magnetic field distributions at birth and the evolutionary pathways of the population. Population synthesis represents a powerful computational framework used to simulate the birth, evolution, and observable properties of the populations of neutron stars in the Galaxy with the aim of comparing with the observed population and constraining physical models. Due to the rich radio pulsar sample, the main efforts of population synthesis studies have been focused on modeling the radio pulsar population alone \citep[e.g.][]{Narayan1990, Faucher2006, Gullon2014, Cieslar2020, Graber2024, Pardo-Araujo2025}. However, radio pulsars represent only a subset of the underlying neutron star population, and analyses based solely on radio detections may bias the inferred birth properties.

Both magnetars and XDINSs probe regimes of magnetic field strengths and evolutionary timescales poorly sampled by the radio pulsar population. Magnetars trace the young high-field end of the birth distribution and exhibit strong magnetic field decay, whereas XDINSs likely represent an evolved population of middle-aged, cooling neutron stars and are believed to be old magnetars \citep{Vigano2013}. Together, these thermally emitting sources offer crucial leverage on the shape and width of the initial magnetic field distribution, on the role of magnetic field decay, and on the connection between different neutron star classes. In particular, understanding the birth magnetic field distribution gives important clues on the formation mechanism during supernova explosions and the fraction of neutron stars that are born as magnetars \citep{Makarenko2021, Pardo-Araujo2026, Shimasue2026}. This in turn has important implications for studying the connections between neutron stars and powerful transients events such as fast radio bursts (FRBs) and gamma-ray bursts (GRBs) \citep[see, e.g.][]{Rea2015, Stratta2018, Beniamini2025, ZhangB2025}. 

Recent advances in magneto-thermal models \citep{Vigano2021, DeGrandis2021, Dehman2023a,Dehman2023b,Ascenzi2024} provide an improved theoretical basis for studying the evolutionary links between these classes of neutron stars and interpreting them in a unified framework. Incorporating realistic prescriptions for the magneto-thermal evolution makes it possible to model not only radio pulsars but also X-ray–bright neutron stars within a single evolutionary scheme. Combining such models with population synthesis has been attempted in only a few studies in the past two decades \citep[e.g.][]{Popov2010, Gullon2015}. These works showed that adding thermally emitting neutron stars helps to break degeneracies between simulation parameters and to better constrain their corresponding ranges, especially for those describing the initial magnetic field distribution. In particular, \citet{Gullon2015} showed that a single log-normal distribution for the initial magnetic fields is unlikely to explain both populations of radio pulsars and magnetars and an extra component extending to fields up to $5 \times 10^{14}$ G is required. 

In this work, we use the software {\tt ML-Poppyns}\footnote{\url{https://github.com/ice-csic-astroexotic/ML-Poppyns}} \citep{Ronchi2021, Graber2024, Pardo-Araujo2025} to perform a comprehensive population synthesis study aimed at constraining the birth and evolutionary properties of isolated neutron stars by jointly considering the radio pulsar population and thermally emitting neutron stars, including magnetars and XDINSs. By combining realistic magneto-thermal evolutionary tracks with emission models in the radio and X-ray bands and survey-specific selection biases, we explore whether a unified scenario with a unique set of birth properties can account for the observed diversity of neutron stars. This multi-population, multi-wavelength approach provides a more complete picture of neutron star birth properties and offers new insights into the connections between distinct observational classes.

The paper is structured in the following way: in Section~\ref{sec:observed_sample}, we describe the observed dataset of radio pulsars and X-ray emitting neutron stars used in this work; in Section~\ref{sec:popsyn}, we explain the ingredients of the {\tt ML-Poppyns} population synthesis framework; in Section~\ref{sec:sbi}, we highlight the simulation-based inference algorithm used to perform the parameter inference; in Section~\ref{sec:results}, we describe the results and in Section~\ref{sec:discussion}, we discuss their implication and compare with other works.


\section{Observed neutron star population}
\label{sec:observed_sample}

\subsection{Radio pulsars}

We consider the same observational sample as described in \cite{Pardo-Araujo2025}, consisting of the radio pulsars detected by three major surveys conducted with Murriyang, the Parkes radio telescope: the \acf{PMPS} \citep{Manchester2001, Lorimer2006}, the \acf{SMPS} \citep{Edwards2001, Jacoby2009}, and the low- and mid-latitude \acf{HTRU} surveys \citep{Keith2010}. For the timing properties, i.e., $P$ and $\dot{P}$, we rely on the ATNF Pulsar Catalogue version 2.6.0. For the radio fluxes, we consider the data from the Thousand Pulsar Array (TPA) program \citep{Johnston2020}, which is part of the large survey project MeerTIME on the MeerKAT telescope. TPA provides a consistently observed sample with well-calibrated flux measurements at 1.429 GHz as reported by \cite{Posselt2023}.
The number of pulsars detected by these three surveys are the following: 
\begin{itemize}
    \item \acs{PMPS}: 1045,
    \item \acs{SMPS}: 218,
    \item \acs{HTRU}: 1095.
\end{itemize}
The discrepancy between the number for the \acs{HTRU} survey reported here and in \cite{Pardo-Araujo2025} is due to reprocessing of archival data that leads to the discovery of 58 new isolated pulsars \citep{Sengar2025}.
For more details on the filters applied to construct this observational dataset, we refer to \cite{Pardo-Araujo2025}.

\subsection{Magnetars, XDINSs and other thermally emitting X-ray pulsars}
\label{sec:obs_x_ns}

\begin{table*}
\caption{Neutron stars with a significant thermal component in the quiescence phase used in this work, classified as magnetars and XDINSs. We report their spin period, $P$, spin-period derivative, $\dot{P}$, and the absorbed flux, $S_{\rm X, abs}$, in the energy range 0.1 - 10 keV. Fluxes values have to be considered with a 10\% relative error.}    
\label{tab:observed_sample_xray}      
\centering   
\begin{tabular}{ccccc}  
\hline\hline       
Source & $P$ [s] & $\dot{P}$ [10$^{-11}$ s s$^{-1}$] & $S_{\rm X, abs}$ [erg s$^{-1}$ cm$^{-2}$] & class \\
\hline 
SGR1627-41 & 2.59 & 1.9 & $4.20 \times 10^{-14}$ & magnetar \\
1E2259+586 & 6.98 & 0.048 & $3.55 \times 10^{-11}$ & magnetar \\
XTEJ1810-197 & 5.54 & 0.283 & $5.30 \times 10^{-13}$ & magnetar \\
SGR1806-20 & 7.75 & 7.5 & $5.49 \times 10^{-12}$ & magnetar \\
CXOUJ1647-4552 & 10.61 & 0.097 & $8.00 \times 10^{-13}$ & magnetar \\
SGRJ0501+4516 & 5.76 & 0.594 & $2.50 \times 10^{-12}$ & magnetar \\
1E1547-5408 & 2.07 & 4.77 & $3.20 \times 10^{-13}$ & magnetar \\
SGRJ0418+5729 & 9.08 & 0.0004 & $1.00 \times 10^{-14}$ & magnetar \\
SGRJ1833-0832 & 7.57 & 0.35 & $6.00 \times 10^{-14}$ & magnetar \\
SwiftJ1822.3-1606 & 8.44 & 0.013 & $2.30 \times 10^{-13}$ & magnetar \\
SwiftJ1834.9-0846 & 2.48 & 0.806 & $1.00 \times 10^{-14}$ & magnetar \\
1E1048.1-5937 & 6.46 & 2.18 & $5.56 \times 10^{-12}$ & magnetar \\
SGRJ1745-2900 & 3.76 & 3.06 & $2.00 \times 10^{-14}$ & magnetar \\
SGRJ1935+2154 & 3.24 & 1.43 & $8.60 \times 10^{-13}$ & magnetar \\
1E1841-045 & 11.79 & 4.09 & $2.33 \times 10^{-11}$ & magnetar \\
SGR1900+14 & 5.20 & 9.2 & $3.92 \times 10^{-12}$ & magnetar \\
4U0142+614 & 8.69 & 0.2 & $1.13 \times 10^{-10}$ & magnetar \\
1RXSJ170849.0-4009 & 11.01 & 1.95 & $3.75 \times 10^{-11}$ & magnetar \\
CXOUJ171405.7-3810 & 3.83 & 6.4 & $1.69 \times 10^{-12}$ & magnetar \\
PSRJ1119-6127 & 0.407 & 0.4 & $4.80 \times 10^{-14}$ & magnetar \\
PSRJ1622-4950 & 4.326 & 1.7 & $9.00 \times 10^{-15}$ & magnetar \\
SwiftJ1818.0-1607 & 1.36 & 5.0 & $2.51 \times 10^{-14}$ & magnetar \\
3XMMJ1852+0033 & 11.559 & 0.014 & $4.46 \times 10^{-13}$ & magnetar \\
\hline
RXJ0420.0-5022 & 3.45 & 0.002759 & $5.01 \times 10^{-13}$ & XDINS \\
RXJ1856.5-3754 & 7.055 & 0.003 & $2.00 \times 10^{-11}$ & XDINS \\
RXJ2143.0+0654 & 9.428 & 0.0041 & $2.51 \times 10^{-12}$ & XDINS \\
RXJ0720.4-3125 & 8.391 & 0.006983 & $1.00 \times 10^{-11}$ & XDINS \\
RXJ0806.4-4123 & 11.37 & 0.0055 & $2.51 \times 10^{-12}$ & XDINS \\
RXJ1308.6+2127 & 10.31 & 0.011 & $3.16 \times 10^{-12}$ & XDINS \\
RXJ1605.3+3249 & 3.39 & - & $7.94 \times 10^{-12}$ & XDINS \\
\hline\hline
\end{tabular}
\end{table*}

The observational sample we consider includes all neutron stars that have a statistically significant thermal component in the soft X-ray band in the energy range 0.1 - 10 keV (see Table~\ref{tab:observed_sample_xray}). The origin of this thermal emission is attributed to the residual heat stored in the neutron star interior and to the Ohmic dissipation of the magnetic field in the neutron star crust \citep{Vigano2013, Vigano2021}. We only consider archival {\it Chandra} and/or {\it XMM-Newton} observations, as they provide the best combination of effective area and angular resolution among past and present X-ray observatories. For all sources, we estimate the absorbed flux in the energy range 0.1 - 10 keV. More details on the data reduction and analysis can be found in \cite{CotiZelati2018, Marino2024, DehmanInPrep}. The resulting observational dataset contains 23 magnetars and 7 XDINSs (one of which is missing the measurement of the spin period derivative, $\dot{P}$). 

Magnetars have been mainly detected through episodes of very energetic outburst emission. However, here we only consider data during their quiescence phase where their emission is mostly of thermal origin and can be described by one or more black-body components possibly with a power-law tail at higher energies. The latter has been attributed to the resonant cyclotron scattering of thermal seed photons emitted from the stellar surface as they interact with the charged particles gyrating around the magnetic field lines in the magnetosphere \citep[see Section~\ref{sec:xray_emission}, ][]{Rea2008, Zane2009, Beloborodov2013}. 

We do not include CCOs in this work since our simulation pipeline does not explicitly account for fallback accretion that could lead to the formation of objects like CCOs \citep{Vigano2012}.

Finally, we note that also RPPs can manifest X-ray spectra with both a thermal and a non-thermal component. For the youngest sources, these two components are usually attributed to the release of heat from the surface and to the synchro-curvature emission from charged particles in the magnetosphere, respectively \citep{Becker1997, Xu2025}. In this work, we exclude RPPs as many of them have spectra contaminated by the contribution of a pulsar-wind nebula \citep[e.g.,][]{Gotthelf2003, Cheng2004}. As the nebula is brighter in X-rays than the central neutron star, this introduces a detection bias which we are not modeling in our simulations. Moreover, contrary to magnetars and XDINSs, most RPPs are already detected in radio and probe a range of magnetic fields which is already well-represented by the rich sample of radio pulsars.


\section{Population synthesis}
\label{sec:popsyn}

To generate a synthetic population of neutron stars, we perform Monte Carlo simulations to model both the dynamical and magneto-rotational evolution of neutron stars. We then model their radio and X-ray emission and apply observational biases both for the radio and X-ray surveys to compare the simulated populations with observations. We follow the same approach as in \citet{Ronchi2021, Graber2024, Pardo-Araujo2025}, and only provide a brief summary of the ingredients here, referring the reader to these works for more detail on the employed methodology. 

\subsection{Dynamical evolution}
\label{subsec:dynamical_evolution}

Following the same approach as in \citet{Pardo-Araujo2025}, we assume the dynamical properties are decoupled from the magneto-rotational ones, allowing us to create a database containing the information of a dynamically evolved population of neutron stars. For this purpose, we simulate a population of $2 \times 10^7$ neutron stars, assigning them uniformly distributed random ages up to $10^8$ yr to ensure a sufficiently large database for the subsequent steps.

For the initial positions, we use the same strategy as outlined in \citet{Ronchi2021}. We assume that the progenitor OB stars follow the Milky Way's spiral arms parametrized by a logarithmic model as in \citet{Yao2017} and the Galactocentric exponential radial distribution found by \citet{Verberne2021} for supernova remnants. We also take into account the local arm \citep[see][]{Ronchi2021}, which is important to model the Sun's neighborhood, especially when reproducing the XDINS population. For the Galactic height, we assume an exponential disk profile \citep{Wainscoat1992} with a characteristic scale height of $\unit[0.18]{kpc}$, consistent with the vertical distribution of young massive stars in the Milky Way \citep{Li2019}. The kick velocities are drawn from the log-normal distribution found in \citet{Disberg2025} with mean of 5.6 and standard deviation of 0.68. 
After setting the initial conditions, we evolve the neutron stars' positions and velocities in time by solving the Newtonian equations of motion with the same prescription as in \cite{Graber2024, Pardo-Araujo2025}. In this way, we obtain a dynamically evolved database of neutron stars that we can sample from to perform the following steps of the simulation, i.e., the magneto-rotational evolution and the detection.

\subsection{Magneto-rotational evolution}
\label{subsec:magneto-rotational_evolution}

In order to model the magneto-rotational evolution of neutron stars, we assume that the initial spin periods follow a log-normal distribution of the form:
\begin{align}
  	\mathcal{P}(\log P_0) &= \frac{1}{\sqrt{2 \pi} \sigma_{\log P}}
      		\, \exp\left(-\frac{(\log P_0 - \mu_{\log P})^2}{2 \sigma_{\log P}^2} \right).
			\label{eqn:P_pdf}
\end{align}
Here, and in the following, we use the subscript ``$\log$'' to refer to $\log_{10}$ to not clutter the notation.

For the initial magnetic field distribution, we consider a double log-normal parametrized in the following way:
\begin{equation}
    \begin{aligned}
      	\mathcal{P}(\log B_0) &= w_{\log B} \mathcal{N}(\log B_0, \mu_{\log B,1}, \sigma_{\log B,1}) \\
        &+ (1-w_{\log B})\mathcal{N}(\log B_0, \mu_{\log B,2}, \sigma_{\log B,2}),
    			\label{eqn:B_pdf_double_lognorm}
    \end{aligned}
\end{equation}
where
\begin{equation}
    \begin{aligned}
      	\mathcal{N}(\log B_0, &\mu_{\log B,i}, \sigma_{\log B,i}) = \\ &\frac{1}{\sqrt{2 \pi} \sigma_{\log B,i}}
          		\, \exp\left(-\frac{(\log B_0 - \mu_{\log B,i})^2}{2 \sigma_{\log B,i}^2} \right),
    			\label{eqn:B_pdf_comp}
    \end{aligned}
\end{equation}
with $i = 1, 2$ and $w_{\log B}$ representing a weight parameter for the first component with a range between 0 and 1.

The choice of this model for the initial magnetic field distribution is given by the fact that a distribution extended to stronger magnetic fields is required to explain the population of magnetars \citep[see][]{Popov2010, Gullon2015, Sautron2025}. The presence of two components in the initial magnetic field distribution implicitly assumes that the birth magnetic field could originate either from different progenitors or mechanisms that enhance the magnetic field during the core collapse \citep{Duncan1992, Barrere2022}.

The initial inclination angle $\chi$ between the spin axis and the magnetic field axis is drawn randomly from a uniform distribution in spherical coordinates, i.e., with a probability $\mathcal{P}(\chi) = \sin \chi$.

After establishing the initial conditions, the spin period and the inclination angle are evolved in time by solving the coupled differential equations \citep{Spitkovsky2006, Philippov2014}:
\begin{subequations}\label{eq:magrot_evol}
\begin{align}
\dot{P} &= \frac{\pi^2}{c^3} \frac{B^2 R^6}{I P} \left( \kappa_0 + \kappa_1 \sin^2 \chi \right), \label{eq:Pdot} \\ 
\dot{\chi} &= -\frac{\pi^2}{c^3} \frac{B^2 R^6}{I P^2} \left( \kappa_2 \sin\chi \cos\chi \right), \label{eq:chidot}
\end{align}
\end{subequations}
where $c$ is the speed of light, $R = \unit[12.59]{km}$ is the neutron star radius for a fiducial neutron star mass $M = \unit[1.4]{M_{\odot}}$ assuming the equation of state BSk24 \citep{Pearson2018}, and $I \simeq 2 M R^2 / 5 \approx \unit[1.78 \times 10^{45}]{g \, cm^{2}}$ is the stellar moment of inertia. For realistic pulsars surrounded by plasma-filled magnetospheres, we choose $\kappa_0 = \kappa_1 = \kappa_2 = 1$.

\subsection{Magneto-thermal evolution models}
\label{subsec:magneto-thermal_models}

\begin{figure}
\includegraphics[width=0.95\columnwidth]{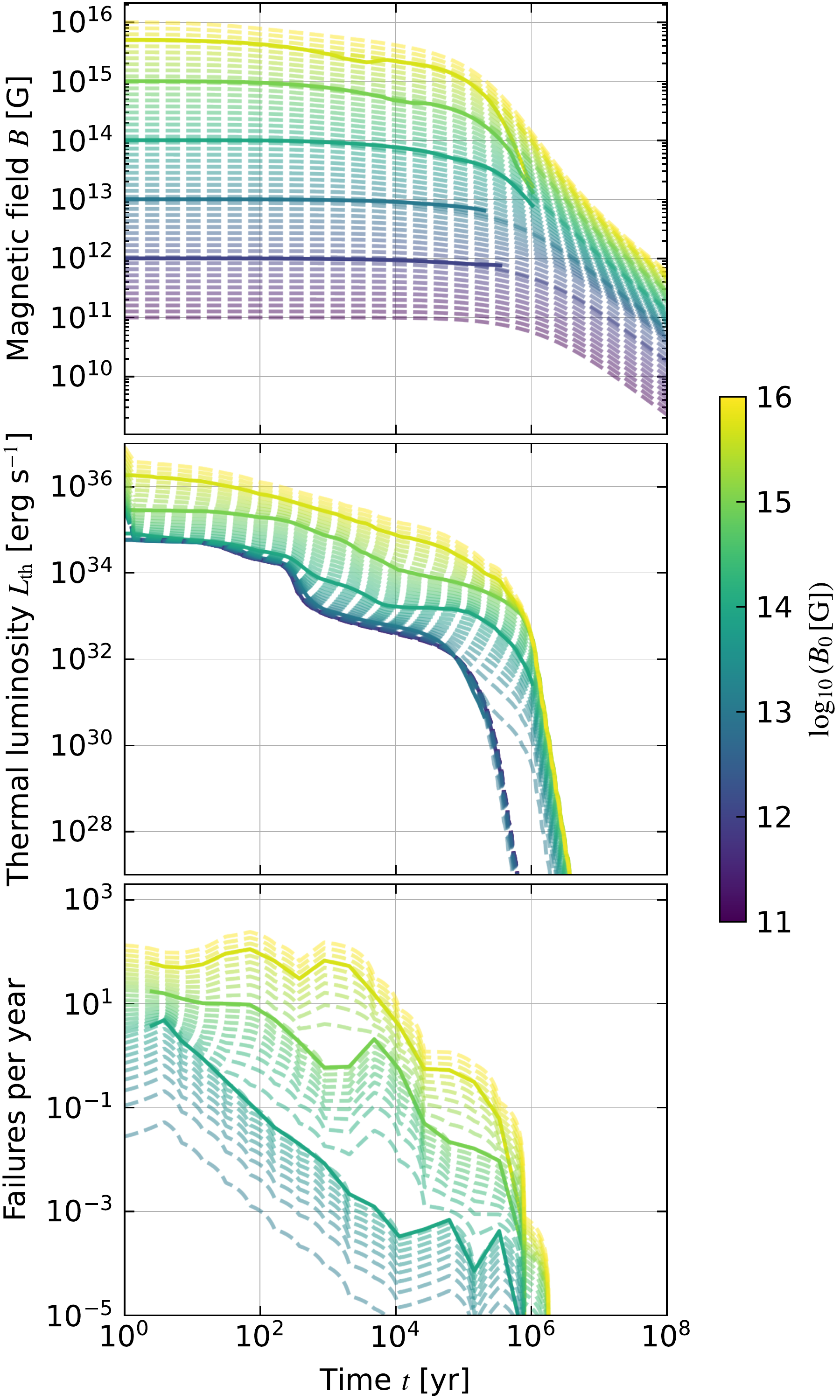}
\caption{\textit{Top panel}: Evolution of poloidal dipolar $B$-field. The five solid curves represent realistic 2D simulations of magneto-thermal evolution in the neutron-star crust \citep{Vigano2021}. The dashed lines represent the evolution predicted by the analytical prescription used in \cite{Graber2024} with the parameter $a_{\rm late} = -1.0$. \textit{Middle panel:} The corresponding evolution of the surface thermal luminosity. The five solid curves represent realistic 2D simulations of magneto-thermal evolution in the neutron-star crust \citep{Vigano2021}. The dashed lines represents the evolution predicted by interpolating the original curve with a bivariate spline and extrapolating to a range of initial dipolar magnetic field values. For $t>10^6$\,yr, we assume a power-law evolution $\propto t^{-10}$ (see text). \textit{Bottom panel:} The corresponding evolution of the rate of crust failures due to magnetic stresses. The three solid curves represent the predicted crust-failure rate from realistic 2D simulations of magneto-thermal evolution in the neutron-star crust \citep{Vigano2021}. The dashed lines represents the evolution predicted by interpolating the original curves with a bivariate spline and extrapolating to a range of initial dipolar magnetic field values. In all panels, the colors represent the initial poloidal dipolar magnetic-field strength, $B_0$.}
\label{fig:B_Lx_magneto-thermal}
\end{figure}

To model the coupled evolution of magnetic field and thermal luminosity of neutron stars, we rely on the results of 2D magneto-thermal simulations \citep[see][]{Vigano2021}. We assume the equation of state BSk24 \citep{Pearson2018}, which has been shown to be able to explain the luminosities of thermally emitting neutron stars \citep{Marino2024}, a neutron star mass $M_{\rm NS} = 1.4$ \msun and a corresponding radius of $R_{\rm NS} = \unit[12.59]{km}$. The impurity parameter in the pasta layer is fixed to 100 \citep{Pons2013}. For the impurity in the outer and inner crust (excluding the pasta layer), the fits of \cite{Carreau2020} have been used (see Fig. 5 in that paper). The blanketing envelope model we adopt is that of \cite{Potekhin2015}, which is composed of heavy elements such as iron and also accounts for magnetic-field effects. For a discussion of the impact of different envelope compositions and the magnetization of the envelope on magneto-thermal evolution, we refer the reader to \cite{Dehman2023c}. Moreover, we consider a crust-confined magnetic field configuration that includes only the dipolar ($\ell = 1$) poloidal and toroidal magnetic field components. We set the two components to have the same maximum magnetic strength. As a result, the poloidal dipole component contains $\sim 90\%$ of the total magnetic energy, while the remaining $\sim 10\%$ are stored in the toroidal component. In the top panel of Fig.~\ref{fig:B_Lx_magneto-thermal}, we show the poloidal dipolar magnetic field evolution with an analytical parametrization for the magnetic field evolution (dashed lines). As current magneto-thermal simulations can only properly model the surface temperature and field evolution until $\sim10^6$ yrs, in this work we use the same parametrization as described in \cite{Graber2024}. This approach captures the trend of field decay for different initial magnetic field strengths by combining several broken power laws together with a late-time power-law evolution with power-law index $a_{\rm late}$. 
Furthermore, to avoid the field decaying to unrealistically small numbers at very late times, we sample the $\log_{10}$ of the final fields from a Gaussian distribution with mean $\mu = 8.5$ and standard deviation $\sigma = 0.5$ in line with the field distribution seen for old millisecond pulsars \citep[see Appendix A in][for more details]{Graber2024}. This allows us to easily extract the dipolar field strength, $B$, at different times, $t$, to compute the magneto-rotational evolution of our synthetic pulsars.

Together with the magnetic field decay, magneto-thermal simulations also provide the coupled evolution of the temperature profile, $T(\theta, t)$, on the neutron star surface, where $\theta$ denotes the polar angular coordinate, in time. Given a temperature profile, $T(\theta, t)$, at a given time, we can compute the total luminosity that a distant observer would estimate by assuming black-body emission from every element of surface area and integrating over the neutron star surface:
\begin{align} \label{eq:thermal_lum}
    L_{\rm th}(t) = 2 \pi \int_0^{\pi} \sigma_{\rm SB} T_{\infty}(\theta, t)^4 R_{\infty}^2 \sin{\theta} {\rm d}\theta.
\end{align}
Here, $\sigma_{\rm SB} = \unit[5.67 \times 10^{-5}]{erg \, cm^{-2} \, K^{-4} \, s^{-1}}$ is the Stefan-Boltzmann constant, and $T_{\infty}$ and $R_{\infty}$ denote the temperature and neutron star radius, respectively, that a distant observer would measure due to the curvature of space-time around the neutron star. These are defined as:
\begin{align}
    T_{\infty} &=  \left( 1 - \frac{2 G M_{\rm NS}}{c^2 R_{\rm NS}} \right)^{1/2} T, \\
    R_{\infty} &=  \left( 1 - \frac{2 G M_{\rm NS}}{c^2 R_{\rm NS}} \right)^{-1/2} R_{\rm NS}.
\end{align}

When the temperature becomes too low, the routines that model the microphysics in magneto-thermal simulations no longer provide reliable results. This happens at ages between $\sim \unit[10^5-10^6]{yr}$. To perform population synthesis simulations at late times, we thus need to extend the luminosity after $\sim 10^6$ yr by considering the cooling during the photon dominated era. In order to get an estimate of the thermal evolution at the late stage, we assume that the surface temperature of the envelope, $T_{\rm e}$, is linked to the temperature at the bottom of the envelope, $T_{\rm b}$, at the interface with the crust via the relation $T_{\rm e} \sim T_{\rm b}^{0.5 + \alpha}$, where $\alpha \sim 0.05$ \citep[see Eq. (34) in][]{Page2004}. During the photon cooling era, we have that $T_{\rm e} \propto t^{-\frac{1}{8\alpha}}$ \citep[see Eq. (42) in][]{Page2004}, which translates into a luminosity evolution $L_{\gamma} \propto T_{\rm e}^4 \propto t^{-\frac{1}{2\alpha}} = t^{-10}$ \citep[see Eq. (38) in][]{Page2004}. Therefore, for the late time evolution of the luminosity, we consider a power-law with the trend $L_{\rm th} \propto t^{-10}$.
In the middle panel of Fig.~\ref{fig:B_Lx_magneto-thermal}, we show the resulting surface thermal luminosity evolution. As we perform simulations for discrete values of the initial poloidal dipolar magnetic fields only, we interpolate using a bivariate spline in order to obtain an estimate of the luminosity for any initial dipolar magnetic field. This approximation is used to extrapolate the luminosity to all initial dipolar magnetic field values in the range $\unit[10^{11} - 10^{16}]{G}$. These interpolated curves allow us to extract the thermal luminosity of a neutron star given its initial poloidal dipolar magnetic field strength and its age.

\subsection{Modeling the X-ray emission}
\label{sec:xray_emission}

Knowing the thermal luminosity at the surface, we can compute an average effective surface temperature as measured by a distant observer:
\begin{align} 
    \bar{T}_{\infty} = \left( \frac{L_{\rm th}}{4 \pi R_{\infty}^2 \sigma_{\rm SB}} \right)^{1/4}.
\end{align}
This temperature can be used to define a black-body spectrum.

Magnetar spectra also show a non-thermal tail that is attributed to resonant cyclotron scattering \citep{Lyutikov2006, Rea2008, Nobili2008}. In this process, seed thermal photons from the neutron star surface interact with the magnetospheric plasma composed mainly of electrons and positrons gyrating along the magnetic field lines. Photons with frequencies, $\omega$, close to the cyclotron frequency, $\omega_{B} = e B / (m_e c)$, will experience resonant scattering boosting their energy. Multiple scattering events distort the black-body spectrum and introduce a power-law tail at high energies. 

To compute the intensity spectrum distorted by resonant cyclotron scattering, $I_{\rm RCS}(E)$, where $E = \hbar \omega$ is the energy of the photons, we use the simplified semi-analytical 1D model described in \cite{Lyutikov2006} (see Appendix~\ref{app:rcs} for more details).

Once the scattered spectrum has been computed, we estimate the total luminosity through the integral:
\begin{align} 
    L_{\rm X}(E) &= 4 \pi R_{\infty}^2 \int_{0}^{2 \pi} \int_{0}^{\pi/2} I_{\rm RCS}(E) \cos{\theta'} \sin{\theta'} {\rm d} \theta' {\rm d}\phi' \\ \nonumber
                 &= 4 \pi R_{\infty}^2 \pi I_{\rm RCS}(E),
\end{align}
where $\theta'$ denotes the angle between the normal to the neutron star surface and a random line of sight direction and $\phi'$ is the azimuthal angle. By knowing the distance $d$ from the neutron star, we can then derive the intrinsic flux density as:
\begin{align} \label{eq:intrinsic_flux}
    S_{\rm X}(E) = \frac{L_{\rm X}(E)}{4 \pi d^2} = \pi \left( \frac{R_{\infty}}{d} \right)^2 I_{\rm RCS}(E).
\end{align}

In order to compute the X-ray flux reaching the Earth, we need to correct for the interstellar absorption due to photoionization. Given an intrinsic (unabsorbed) X-ray flux density, $S_{\rm X}(E)$, where $E$ denotes the energy in keV, the absorbed flux can be derived by \citep[see][]{Wilms2000}:
\begin{align} \label{eq:absorbed_flux}
    S_{\rm X, abs}(E) = e^{-\sigma_{\rm ISM}(E) N_{\rm H}} S_{\rm X}(E).
\end{align}
Here, $\sigma_{\rm ISM}(E)$ in units of cm$^2$ per hydrogen atom represents the energy-dependent effective absorption cross section of the \acf{ISM}, which is normalized to the hydrogen atom number. Taking into account the phases of the \acs{ISM}, this cross section can be written as:
\begin{align} 
    \sigma_{\rm ISM}(E) = \sigma_{\rm gas}(E) + \sigma_{\rm molecules}(E) + \sigma_{\rm grains}(E).
\end{align}
To compute $\sigma_{\rm ISM}(E)$, we will only take into account the neutral atomic gas, that is the $\sigma_{\rm gas}(E)$ term, neglecting the possibility of ionization and the presence of molecules and grains in the \acs{ISM}. As shown in \citet{Wilms2000} \citep[see also][]{Willingale2013}, including these effects would give only minor corrections to the total effective cross sections.
Therefore, we can rewrite the equation above as \citep[see][]{Wilms2000}:
\begin{align} \label{eq:ism_cross_section}
    \sigma_{\rm ISM}(E) \sim \sigma_{\rm gas}(E) \sim \sum_Z A_Z \sigma_{{\rm ph,} Z}(E),
\end{align}
where, for an element with atomic number $Z$, the relative abundance with respect to hydrogen is defined as $A_Z = N_Z / N_{\rm H}$. Table 2 in \cite{Wilms2000} reports the logarithm of the relative abundance $\log_{10} \left( N_Z / N_{\rm H} \right) + 12$, where, by definition, the abundance of hydrogen is set to 12. Moreover, $\sigma_{{\rm ph,} Z}(E)$ denotes the photoionization cross section. To derive the total effective cross section, $\sigma_{\rm ISM}(E)$, we have translated the \texttt{Fortran 77} routines written by \citet{Balucinska-Church1992} into \texttt{Python}. The original routines implement polynomial fits of the atomic photoionization cross sections in the energy range of \unit[0.03 -- 10]{keV} for seventeen elements.

In Eq.~\eqref{eq:absorbed_flux}, $N_{\rm H}$ denotes the hydrogen column density, which has units of hydrogen atoms\,cm$^{-2}$ and is defined as the integral of the spatial density of neutral hydrogen, $n_{\rm H}$, along the line of sight:
\begin{align}
    N_{\rm H} = \int n_{\rm H} {\rm d}l,
\end{align}
where $l$ represents the path length along the line of sight. For a given neutron star with known equatorial sky coordinates (RA, DEC) and distance, $d$, the value of $N_{\rm H}$ can be estimated using the reddening map of the Galaxy and the calibration factor provided by \citet{Doroshenko2024}. This factor converts reddening, $E(B-V)$, provided by the map into an $N_{\rm H}$ value assuming a set of abundances for the \acs{ISM}. By default we use the abundances specified in \citet{Wilms2000}. 

Finally, to compute the total flux, we integrate the flux density in Eq.~\eqref{eq:absorbed_flux} in the energy band $0.1 - 10$\,keV:
\begin{align} \label{eq:absorbed_flux_int}
    S_{\rm X, abs} = \int_{0.1}^{10} S_{\rm X, abs}(E) {\rm d}E.
\end{align}

\subsection{Crustal failures and outburst activity}
\label{sec:crust_failures}

Magnetars have been primarily discovered through their flaring and bursting activity. This is often accompanied by sudden increases in emission by several orders of magnitude with respect to their persistent thermal and non-thermal emission that could last for months or years, a phenomenon commonly referred to as an outburst \citep{Rea2011, CotiZelati2018}. This highly variable and powerful activity is likely linked to reconnection events in neutron stars magnetosphere, possibly triggered by the star’s interior evolution. In particular, current models suggest that the evolution and instability of ultra-strong magnetar magnetic fields causes mechanical stresses in the neutron star crust. These accumulate to the point of exceeding the breaking strain of the crustal lattice, generating an event that can locally heat the matter and drag magnetic field lines along. The resulting perturbation can then propagate outwards triggering magnetospheric activity \citep{Beloborodov2009, Chugunov2010, Carrasco2019, Dehman2020, Lander2023}.
As the flaring and outburst activity represents a crucial bias towards the discovery of highly magnetized neutron stars, we need to account for this in our model and predict the rate of these events for a given star. To this end, 2D magneto-thermal simulations also provide an estimate of when and where the crust fails due to the aforementioned stresses \citep{Perna2011,Dehman2020}. This information can then be used to compute the rate of crustal failures as a function of time and initial poloidal dipolar magnetic field strength during the magnetic field evolution. To evaluate the rate, we divide the total evolution time span into logarithmic bins and count how many failure events are produced in each bin. To obtain the rate of failures in each bin, we then divided the total count in each bin by the bin width.

In the bottom panel of Fig.~\ref{fig:B_Lx_magneto-thermal}, we show the rate of crust failures as a function of time predicted by our 2D simulations of magneto-thermal evolution. As we perform simulations for discrete values of the initial dipolar magnetic field only, we interpolate using a bivariate spline to obtain an estimate of the failure rate for any initial dipolar magnetic field. This approximation is used to extrapolate the rate to all initial dipolar magnetic field values in the range $\unit[10^{11} - 10^{16}]{G}$. As for the thermal luminosity, these interpolated curves allow us to obtain the crust-failure rate of a neutron star given its initial poloidal dipolar magnetic field strength and age.

\subsection{Radio detection}

To compute the radio fluxes, beam geometry and propagation effects due to the interaction of the radio waves with the ISM, we use the same prescription as in \cite{Graber2024, Pardo-Araujo2025}. In particular, to model the intrinsic radio luminosity $L_{\rm int}$, we adopt Eq. (7) in \citet{Pardo-Araujo2025} where the mean of the logarithm of the radio luminosity, $\mu_{\log L_0}$, and the power-law index, $\alpha_{L}$, determining the dependence on the spin-down power, are free parameters. 
For the radio emission geometry, we consider two filled cone-shaped emission beams centered at the two magnetic poles with a period-dependent half opening angle $\rho_{\rm b} \propto P^{-1/2}$ \citep[see Eq. (13) in][]{Graber2024}.
A pulsar can be detected if its radio beam crosses our line of sight, whose direction relative to the pulsar spin axis is assumed to be random. In this case, the intrinsic bolometric radio flux can be computed as:
\begin{equation}
	S = \frac{L_{\rm int}}{\Omega_{\rm b} d^2}, 
\end{equation}
where $\Omega_{\rm b}$ is the solid angle covered by the two radio beams and $d$ is the distance of the pulsar from Earth. To determine the radio flux density, $S_{f}$, (measured in Jy) at a central observing frequency, $f$, of a specific survey, we assume a power-law spectrum with spectral index of $-1.8$ \citep{Posselt2023}. The flux density, $S_{f , {\rm obs}}$, that reaches Earth is then given by:
\begin{equation}
	S_{f , {\rm obs}} \simeq S_{f} \frac{ w_{\rm int}}{w_{\rm obs}}
	\label{eqn:obs_flux} 
\end{equation}
where $w_{\rm int}$ is the intrinsic pulse width and $w_{\rm obs}$ is the observed pulse width after propagation effects in the ISM have been taken into account \citep[see Eqs.~(15) and (19) in][]{Graber2024}.

As in \citet{Graber2024, Pardo-Araujo2025}, we consider three major radio surveys conducted with Murriyang, the Parkes radio telescope: the \acf{PMPS} \citep{Manchester2001, Lorimer2006}, the \acf{SMPS} \citep{Edwards2001, Jacoby2009}, and the low- and mid-latitude \acf{HTRU} surveys \citep{Keith2010}. A summary of all relevant survey parameters is provided in Table 1 in \citet{Graber2024}. For each simulated neutron star whose radio beam intercepts our line of sight, we compute the signal-to-noise ratio using the radiometer equation \citep{Lorimer2012}:
\begin{equation}
        S/N = \varepsilon \frac{ S_{\rm mean} G \sqrt{n_{\rm pol} \Delta f_{\rm bw} t_{\rm obs}} }
        		{ \beta \left[ T_{\rm sys} + T_{\rm sky }(l, b) \right] } 
		\sqrt{\frac{P- w_{\rm obs}}{w_{\rm obs}}},
		\label{eqn:radiometer}
\end{equation}
where $S_{\rm mean} \simeq S_{f, {\rm obs}} w_{\rm obs} / P$ is the mean flux density.
For a more detailed description of all relevant parameters in this equation, we refer to \citet{Graber2024, Pardo-Araujo2025}.
We only stress here that compared to \citet{Graber2024, Pardo-Araujo2025}, we have updated the radiometer equation by including the efficiency factor, $\varepsilon$, which is a function of the duty cycle, $\delta = w_{\rm obs}/P$, as determined by \citet{Morello2020} (see their Eq. (44)). Since pulsars are predominantly found via incoherent searches based on fast Fourier transforms (FFTs), this efficiency factor models the decay in sensitivity of pulsar searches when looking for pulsars with small duty cycles. Sometimes this efficiency factor is directly incorporated into the degradation factor, $\beta$, by considering values larger than $\sim 1.25$ \citep[see the discussion in][]{Morello2020}. Here, we fix $\beta$ to $1.25$ assuming that it only accounts for imperfections during the digitization of the signal.

\subsection{Modeling the X-ray detection biases}

The observed sample of X-ray emitting neutron stars is difficult to reproduce in a simulation framework as the sample is subject to complex observational biases which are not well under control. Many of these sources have been discovered through targeted observations with different X-ray instruments. Therefore, the sky coverage and corresponding threshold sensitivities are very inhomogeneous and difficult to reproduce in a simulation framework. Indeed, many magnetars have been discovered through their outburst activity whose high energy emission triggered all-sky X-ray monitors onboard Swift--BAT or Fermi--GBM. This has enabled follow-up campaign observations that allowed us to detect a periodicity and identify these sources as neutron stars \citep{Rea2011, CotiZelati2018}.
Moreover, in the observed catalog, for sources that underwent a magnetar-like outburst, we are only considering those with detected quiescent emission (note that some magnetars were identified as such during an outburst event but their quiescent emission is too faint to be detectable). On the other hand, sources like XDINSs have very stable X-ray emission over timescales of several decades, and despite their relatively faint X-ray luminosities can be detected due to their high fluxes, a consequence of their close-by distances. To encompass these different observational biases, we adopt a simplified approach to model the X-ray detection.

First, we consider an all-sky coverage for X-ray surveys, i.e., we assume that X-ray instruments have scanned the entire sky. We assume that for fluxes in the energy range $0.1 - 10$\,keV, the observed sample is complete above $\sim \unit[10^{-12}]{erg \, s^{-1} \, cm^{-2}}$. When representing the observed X-ray population in the $\log N - \log S$ plane, where $N$ represents the number of sources above a given flux value $S$ (see Section~\ref{sec:results}), we observe a change in slope at fluxes lower than $\unit[10^{-12}]{erg \, s^{-1} \, cm^{-2}}$ (see black line in Figs.~\ref{fig:logN-logS_sim_vs_obs} and \ref{fig:logN-logS_sim_vs_obs_youngxdins}), meaning that we are likely missing sources below this flux value. Therefore, we consider an average flux threshold of $\unit[10^{-12}]{erg \, s^{-1} \, cm^{-2}}$ in the X-ray band $\unit[0.1-10]{keV}$ with a standard deviation of 0.5 (in log) in our simulations. This assumption is also in agreement with average flux thresholds of early all-sky surveys as those conducted with ROSAT \citep{Truemper1982, Voges1999}. Ultimately, the implementation of this all-sky threshold allows the detection of all sources that are intrinsically very bright or close-by like the XDINSs.

On top of this first filter, we assume a second filter for all simulated magnetar-like sources that show outburst activity. In particular, for every simulated neutron star, we estimate the current crust-failure rate given its initial magnetic field and age as explained in Section~\ref{sec:crust_failures}. Given this rate, we then estimate the probability of each star to have experienced a crust failure event that could have triggered an outburst in the last 30 years. This value is motivated by the fact that continuous all-sky monitoring capable of detecting new activity from magnetars has only been available for approximately the past 30 years, beginning with the launch of the Rossi X-ray Timing Explorer (RXTE) mission \citep{Bradt1993}. Prior to this, transient magnetar activity had been significantly harder to detect. Moreover, we assume that all outburst events are energetic enough to trigger a detection and deep follow-up observations with X-ray instruments. For the latter, we assume an average flux threshold of $\unit[10^{-14}]{erg \, s^{-1} \, cm^{-2}}$ with a standard deviation in log of 0.5. We assume this flux threshold to match the faintest detected magnetar (i.e., SGRJ0418+5729 in Table~\ref{tab:observed_sample_xray}). 

In summary, the choice of flux threshold values considered in this section comes from both trying to reproduce the observed X-ray flux distribution and the typical sensitivity of X-ray instruments for short and long exposure times \citep[see for example Fig.~3 in][]{Watson2001, Weisskopf2002, Gehrels2004}.

\subsection{Representation of simulated output}
\label{sec:sim_maps}

The output of the simulations, consisting of detected mock neutron stars in radio and X-rays, is represented in the form of 2D $P-\dot{P}$ maps following the same strategy as in \cite{Graber2024, Pardo-Araujo2025}. In particular, for each simulated survey, we produce two maps: a density $P-\dot{P}$ map and an average flux map of the $P-\dot{P}$ diagram. Both types of maps have ranges set to $[0.01, 100] \, {\rm s}$ and $[10^{-20}, 10^{-9}] \, {\rm s \, s^{-1}}$ for the $P$ and $\dot{P}$ axis, respectively, and a resolution of 32 bins along both axes. The density map contains information on the number of pulsars detected in each bin, while the flux maps store the average value of the radio flux in Jy or the X-ray flux in erg s$^{-1}$ cm$^{-2}$ of the neutron stars falling into that specific bin for the radio surveys or X-ray survey, respectively. Therefore, the output of each simulation is summarized by a total of eight 2D maps, six for the radio surveys and two for the X-ray survey. 

\begin{figure*}

\centering
\includegraphics[width = 0.8\textwidth]{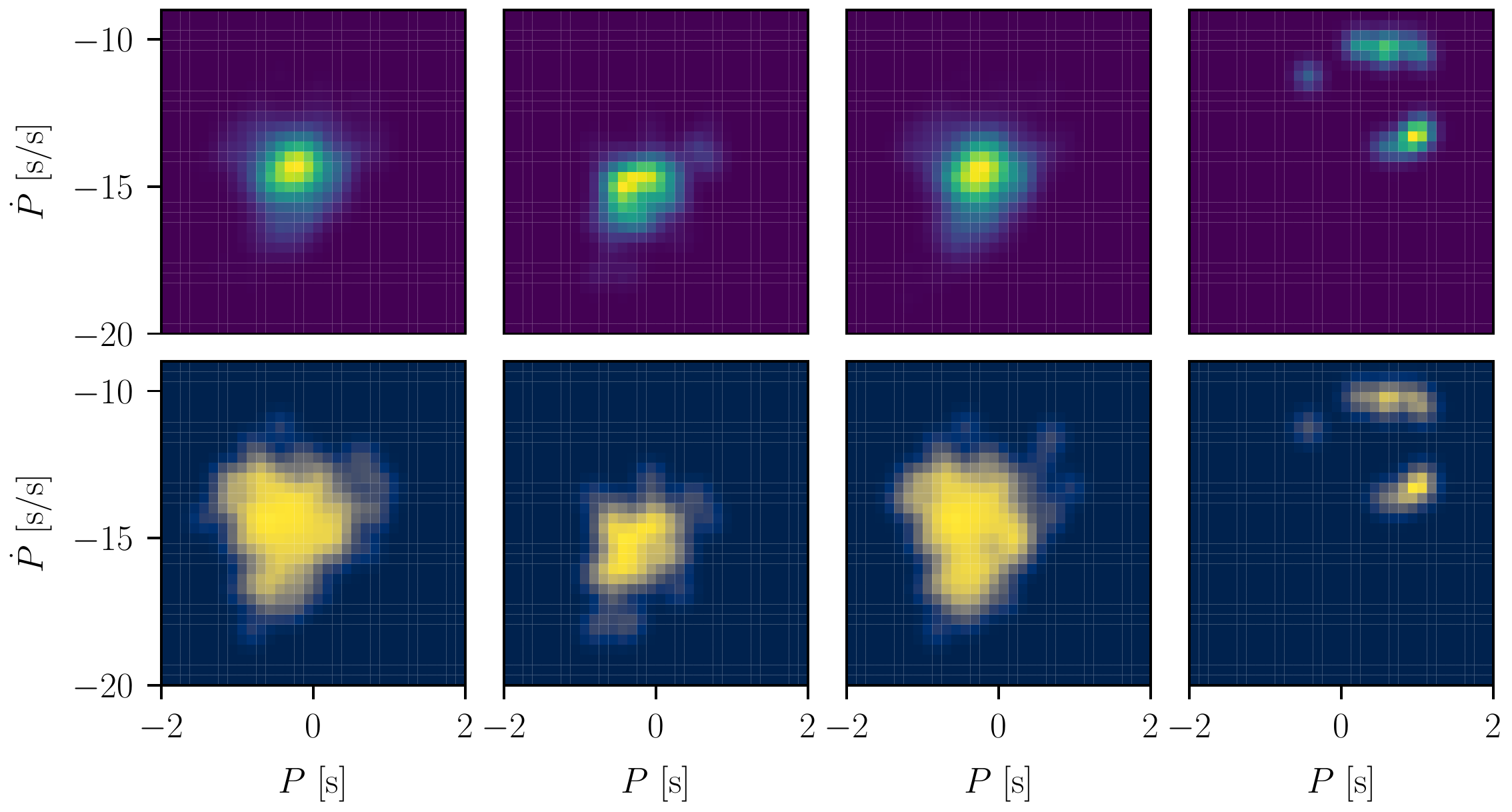}
\caption{The eight density maps for the observed population of neutron stars used in the experiment described in Section~\ref{sec:results_youngxdins}. The top and the bottom row show the $P$-$\dot{P}$ diagrams and the $P$-$\dot{P}$ averaged flux maps, respectively, for the (from left to right) PMPS, SMPS, HTRU, and X-ray surveys. In the top row, the color represents the density in neutron star number within each bin, while in the bottom row the color represents the averaged flux in Jy for the radio survey maps and in erg s$^{-1}$ cm$^{-2}$ for the X-ray survey map within each bin. In the bottom row, for bins without any stars, the averaged flux in log scale has been set to -7 and -17 for the radio and X-ray maps, respectively.}
\label{fig:density_maps_observed}
\end{figure*}

In Fig.~\ref{fig:density_maps_observed}, we show the corresponding maps for the real observed population. Note that for the X-ray population (last maps on the right) the top and bottom cluster represent the young magnetar population (with estimated ages less than 2 kyr, see Section~\ref{sec:results} for more details) and the older XDINS population, respectively.

In order to smooth out abrupt features in the maps due to the random nature of the simulation and stabilize the subsequent machine learning pipeline, we apply a Gaussian smoothing kernel. The resulting maps associated with the respective ground truth labels, i.e., the parameter set used to simulate them, will be the input provided to train the simulation-based inference framework explained in Section~\ref{sec:sbi}. To further stabilize the training procedure, we standardize maps and labels so that the values have a mean of 0 and a standard deviation of 1. Standardization is performed on a sample basis for the maps and on a dataset basis for the labels \citep[see][for more details]{Graber2024, Pardo-Araujo2025}.

\section{Simulation-based inference}
\label{sec:sbi}

In recent years, the development of machine learning has allowed the emergence of new tools to perform parameter inference. Simulation-based inference (SBI) is a recently developed framework that combines the power of Bayesian statistics and deep learning to optimize and perform parameter inference when using complex model simulators \citep[see][for reviews]{Cranmer2020, Zammit-Mangion2025}. In this work, we use a SBI approach called \acf{NPE} \citep[e.g.][]{Papamakarios2016, Lueckmann2017, Greenberg2019, Dax2021, Mishra-Sharma2022, Vasist2023, Barret2024}, which has been adopted to infer neutron star population parameters in recent population synthesis studies \citep{Graber2024, Pardo-Araujo2025, Sautron2025}. We refer the reader to these works for a more complete overview and description of the different \acs{SBI} methods and of \acs{NPE} in particular. We only summarize the main points here. 

In \acs{NPE}, a neural density estimator is trained to directly map a simulation output, $\bx$, to the posterior distribution, $\pr(\bt | \bx)$. This gives the probability for a set of model parameters, $\bt$, to have generated the output, $\bx$, through a stochastic simulator. The neural density estimator, $q$, is parametrised by a neural network, $F$, with weights, $\phi$, i.e., $q_{F(\bx, \boldsymbol{\phi})}$. The network is then optimised by minimizing the following loss function
\begin{equation}
	\mathcal{L}(\boldsymbol{\phi}) = - \sum_{i=1}^M \ln q_{F(\bx_i, \boldsymbol{\phi})} (\bt_i)
		\label{eqn:loss}
\end{equation}
over a training data set $\{\bt_i, \bx_i\}$ with a  total of $M$ simulations. This loss is minimized when the neural density estimator approximates the true posterior, that is when
\begin{align}
    q_{F(\bx, \boldsymbol{\phi})}(\bt) \approx \pr (\bt | \bx).
\end{align}
In this work, we adopt a sequential version of \acs{NPE} called \acf{TSNPE} \citep{Deistler2022}. The workflow of this algorithm is as follows \citep{Pardo-Araujo2025}:
\begin{enumerate}
    \item Sample the proposal prior distribution to obtain $\bt_i\sim\pr(\bt)$ for $i=1,...,M$.
    \item Using the simulator, generate synthetic data $\bx_i \sim \pr(\bx|\bt_i)$ based on $\bt_i$ from step 1. 
    \item Train the neural density estimator on the dataset composed of pairs $(\bt_i, \bx_i)$ obtained in the previous steps.
    \item Use the trained neural density estimator to approximate the posterior distribution, $\pr (\bt | \bx_0)$, at the observed data, $\bx_0$.
    \item Restrict the prior distribution to the approximated posterior distribution computed in step 4 \citep[see Section 3.3 in][and references therein]{Pardo-Araujo2025}.
    \item Update the proposal prior distribution with the new restricted prior and return to step 1.
\end{enumerate}
Steps 1 to 6 are performed several times and each iteration is commonly referred to as {\it training round}. This way, the posterior is iteratively refined focusing the computational resources on the region of the parameter space that is more compatible with the observed data. In each round, the newly generated simulations are added to the ones used in the previous rounds, so that the combine simulations are employed together as the training dataset. 

\subsection{Neural posterior estimation setup}
\label{sec:TSNPE_setup}

In this work, we use the same setup as in \cite{Graber2024, Pardo-Araujo2025}, and only summarize the main aspects important for this study.
To set up the inference procedure and the neural density estimator, we use the library \texttt{sbi}\footnote{\url{https://github.com/sbi-dev/sbi}} \citep[v=0.22.0,][]{Tejero-Cantero2020, BoeltsDeistler_sbi_2025}.
The output of the simulations is represented in the form of eight 2D maps as described in Section~\ref{sec:sim_maps} and is processed by a \acf{CNN} as an eight-channel input. The \acs{CNN} architecture is constituted by two 2D-convolutional layers, each followed by a 2D max pooling layer and a ReLU activation function \citep{Glorot2010}, extracting key features into a latent vector of size 32. This latent vector is the input to the neural density estimator which we set to be a \acf{MDN}. Our MDN consists of three fully connected hidden layers with 32 neurons each and ReLU activations. The output layer is divided into four components that predict the mixture weights, means, and the diagonal and upper-triangular elements of the Gaussian covariance matrices \citep[see Fig. 6 in][]{Graber2024}. We use ten Gaussian components in the mixture to ensure sufficient flexibility when approximating the posterior.
We fix the batch size to eight, the fraction of training dataset used for validation to 0.1 and train both the \acs{CNN} and the \acs{MDN} simultaneously using the Adam optimiser \citep{Kingma2014} with an initial learning rate of $5 \times 10^{-4}$ and an early stopping criterion of 20 epochs to prevent overfitting. This implies that training is stopped if the validation metric (also given by Eq.~\eqref{eqn:loss} computed over the validation dataset) does not improve for 20 consecutive epochs, with the best validation weights being saved. At each training round, the weights for the \ac{CNN} are re-initialized using the Kaiming prescription \citep{Kaiming2015}, while for the \ac{MDN} the weights are initialized with PyTorch's default initialization. Moreover, in order to take into account the variability of the initialization and the training process in each round, we train an ensemble of five neural networks with identical architectures as described above. The final posterior in each round is then given as a weighted sum of the individual posterior distributions computed by each network based on equal weights \citep[see][for more details]{Graber2024, Pardo-Araujo2025}. 
In each of the experiments described below in Sections~\ref{sec:results_allpop} and~\ref{sec:results_youngxdins}, we train the TSNPE algorithm over ten rounds. In the first round, we use 30000 simulations for training and validation and 3000 for testing purposes, and adopt the following uniform prior ranges for the parameters:
\begin{align}
\label{eqn:priors_double_lognormal}
    \mu_{\log P} & \in \mathcal{U}(-1.5, 0.5), \nonumber \\
    \sigma_{\log P} &\in \mathcal{U}(0.1, 1),  \nonumber \\
    \mu_{\log B, 1} &\in \mathcal{U}(12, 13.5), \nonumber \\
    \sigma_{\log B, 1} &\in \mathcal{U}(0.1, 1), \nonumber \\
    \mu_{\log B, 2} &\in \mathcal{U}(13.5, 15), \\
    \sigma_{\log B, 2} &\in \mathcal{U}(0.1, 1), \nonumber \\
    w_{\log B} &\in \mathcal{U}(0.1, 1), \nonumber \\
    a_{\rm late} &\in \mathcal{U}(-3, -0.5), \nonumber \\
    \mu_{\log L_0} &\in \mathcal{U}(24.6 , 28.6), \nonumber \\
    \alpha_{L} &\in \mathcal{U}(0.1, 1). \nonumber 
\end{align}

In each of the following rounds, we generate 1000 new simulations by sampling the restricted prior. This number has been chosen arbitrarily but as in \cite{Pardo-Araujo2025}, it guarantees a good compromise between computational costs and network convergence as shown below. We then retrain the ensemble on the new dataset formed by adding the new simulations to the ones of the previous rounds.
The training process is executed on a Tesla V100 SXM2 GPU with $\unit[32]{GB}$ of memory. The generation of simulations in each \ac{TSNPE} round are parallelized for both the training and test datasets to speed up the algorithm. For this, we use the Python package Dask \citep{dask}, a library for dynamic task scheduling. In total, $600$ CPU workers are employed to handle the parallelized simulations.


\section{Results}
\label{sec:results}

\begin{table}
\caption{Comparison between best parameter values inferred by considering the full sample of radio pulsars and X-ray emitting neutron stars (Radio + X-ray) and by considering the radio pulsars and a restricted sample with only the young magnetars and the XDINSs (Radio + young mag, XDINSs). For each parameter, we report the median value together with the 95\% credibility level derived from the posteriors.}    
\label{tab:infer_results}      
\centering  
\begin{tabular}{c c c}  
\hline\hline       
Parameter & Radio + X-ray & Radio + young mag, XDINSs \\
\hline 
$\mu_{\log P}$ & $-0.19^{+0.45}_{-0.73}$ & $-0.26^{+0.44}_{-0.46}$ \\
$\sigma_{\log P}$ & $0.61^{+0.28}_{-0.33}$ & $0.70^{+0.23}_{-0.34}$ \\
$\mu_{\log B, 1}$ & $12.70^{+0.26}_{-0.23}$ & $12.72^{+0.20}_{-0.21}$ \\
$\sigma_{\log B, 1}$ & $0.45^{+0.11}_{-0.12}$ & $0.45^{+0.13}_{-0.13}$ \\
$\mu_{\log B, 2}$ & $13.85^{+0.41}_{-0.32}$ & $14.11^{+0.53}_{-0.51}$ \\
$\sigma_{\log B, 2}$ & $0.35^{+0.18}_{-0.19}$ & $0.49^{+0.35}_{-0.32}$ \\
$w_{\log B}$ & $0.83^{+0.16}_{-0.22}$ & $0.74^{+0.21}_{-0.25}$ \\
$a_{\text{late}}$ & $-0.89^{+0.31}_{-0.32}$ & $-0.85^{+0.26}_{-0.26}$ \\
$\mu_{\log L_0}$ & $25.78^{+0.24}_{-0.24}$ & $25.70^{+0.25}_{-0.26}$ \\
$\alpha_L$ & $0.71^{+0.09}_{-0.09}$ & $0.73^{+0.09}_{-0.09}$ \\ 
\hline\hline
\end{tabular}
\end{table}

\subsection{Inference on the entire population}
\label{sec:results_allpop}


\begin{figure*}
\centering
\includegraphics[width = 0.5\textwidth]{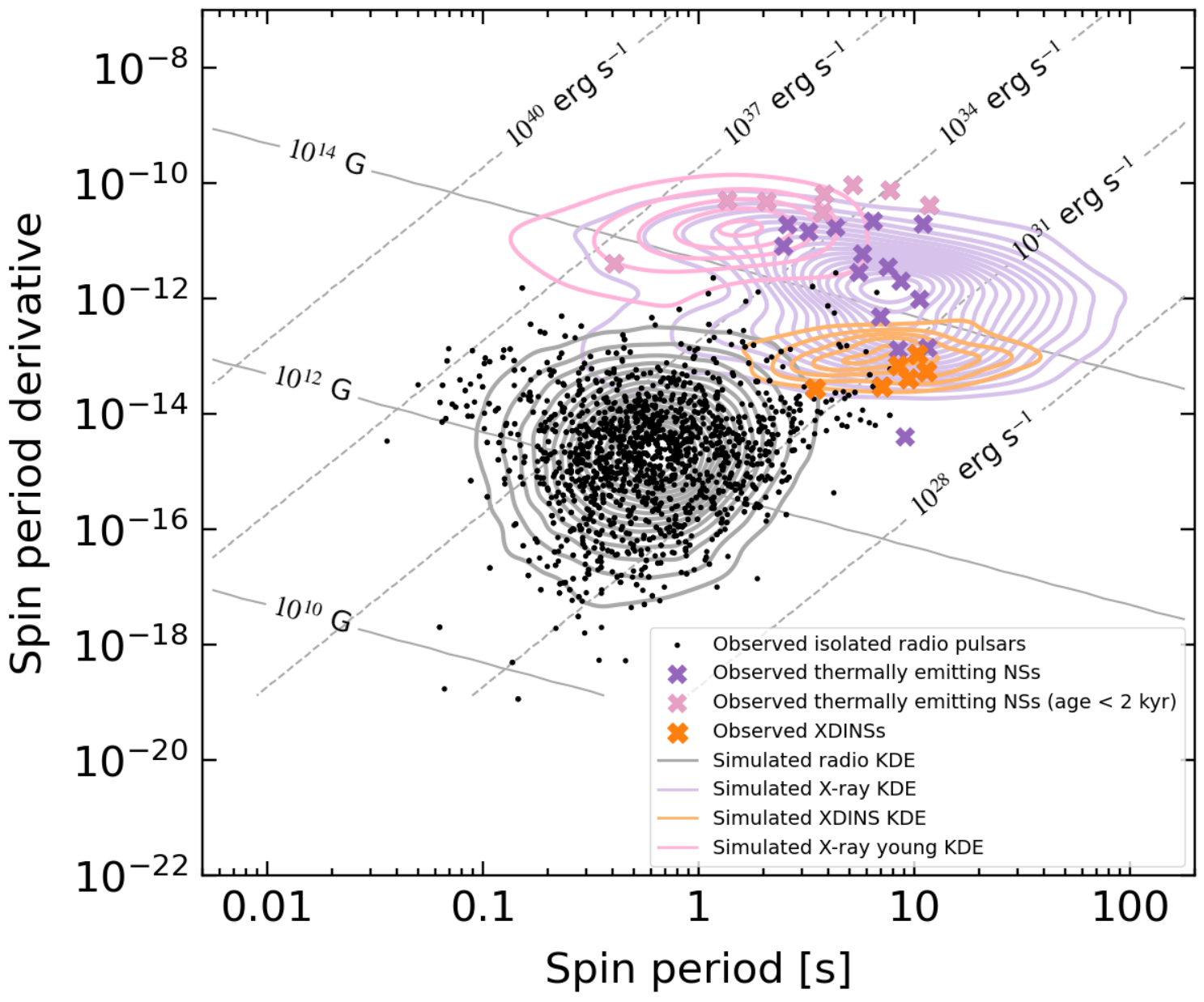}
\caption{Comparison between the observed populations of radio pulsars and X-ray emitting neutron stars and the best simulated populations in the $P-\dot{P}$ plane for the experiment described in Section~\ref{sec:results_allpop}. Points and crosses represent the observed radio pulsars and X-ray emitting neutron stars, respectively. Among the X-ray-emitting neutron stars, pink crosses indicate young magnetars (estimated ages $<2$ kyr), while orange crosses denote the observed XDINSs. The contour lines show KDE density contours derived from 100 simulations with input parameters sampled from the round-5 posterior distribution. The pink and orange contours correspond to the two simulated sub-populations, while the purple contours represent the full simulated X-ray-emitting neutron star population.}
\label{fig:PPdot_sim_vs_obs}
\end{figure*}


\begin{figure*}
\centering
\includegraphics[width = 0.45\textwidth]{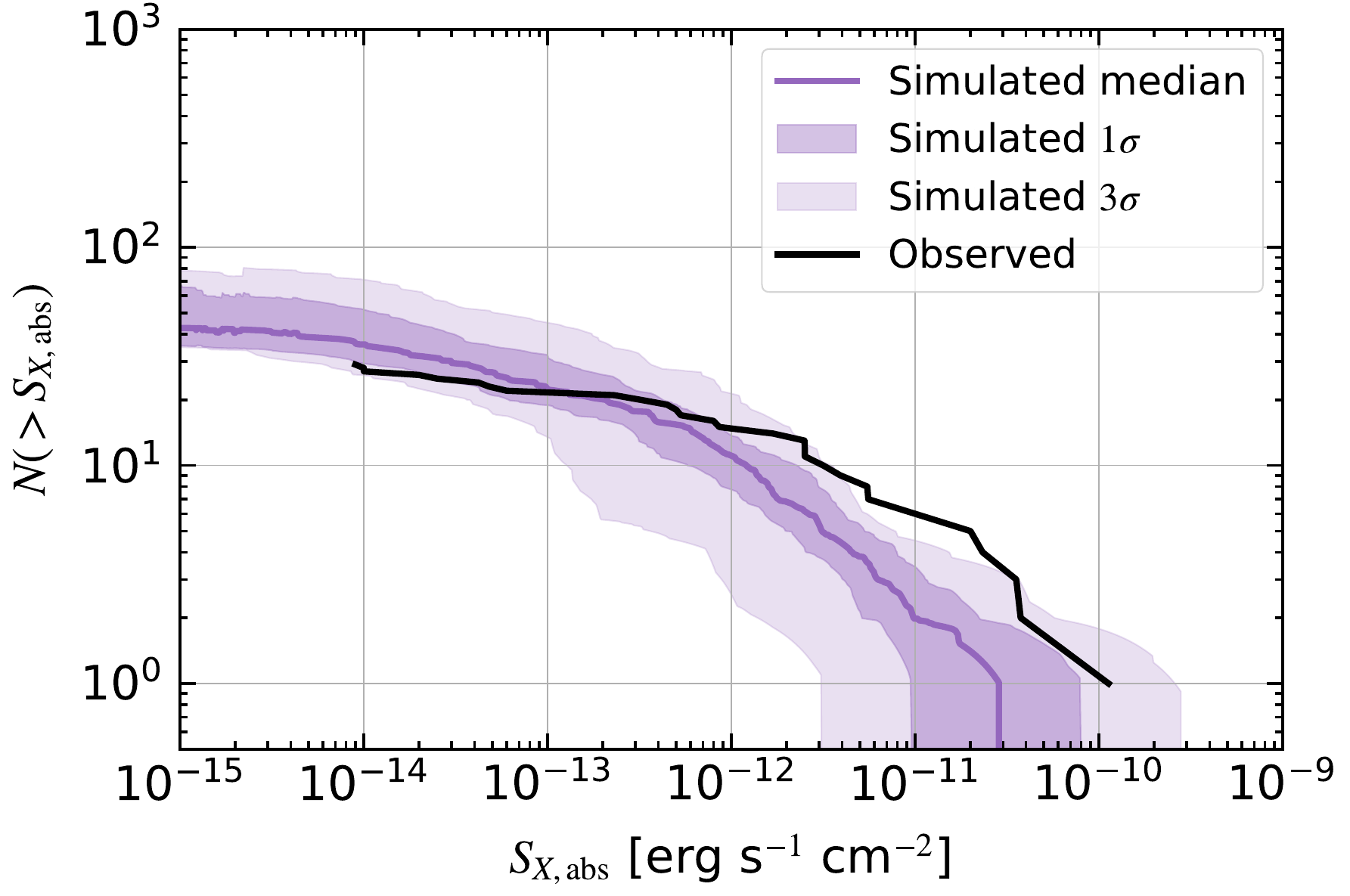}
\caption{Comparison of the $\log N-\log S$ distributions for the observed and simulated thermally emitting X-ray neutron stars for the experiment described in Section~\ref{sec:results_allpop}. The black line represents the trend for the observed population while the purple line and bands represent the median, the 1-$\sigma$ and 3-$\sigma$ uncertainties, respectively. The latter are computed over 100 simulations generated from sets of parameters drawn from the posterior distribution obtained from round 5 of our TSNPE algorithm.}
\label{fig:logN-logS_sim_vs_obs}
\end{figure*}


We first train the TSNPE algorithm over 10 rounds with the setup explained in Section~\ref{sec:TSNPE_setup} considering the entire observed population of radio and X-ray emitting neutron stars described in Section~\ref{sec:observed_sample}. This accounts for 1045, 218, and 1095 radio pulsars in the PMPS, SMPS, and HTRU surveys, respectively, as well as 29 X-ray-emitting neutron stars (i.e., the entire sample listed in Table~\ref{tab:observed_sample_xray}, with the exception of RXJ1605.3+3249, which does not have a $\dot{P}$ measurement).

After round 5 the approximated posterior results are stable (see Appendix~\ref{app:posterior_full_pop}). Therefore, we use the model trained in this round to infer the best parameter values. In the first column of Table~\ref{tab:infer_results}, we list the median of the best inferred parameters from round 5 with the 95\% credible interval. In particular, for the initial magnetic field distribution, we find that the two log-normal components peak at $5\times 10^{12}$ G and $7.1\times 10^{13}$ G, respectively. To assess the quality of the trained posterior estimator in this round, we check its predictive power over the test dataset by comparing its posterior predictions with the ground truth labels (see Fig.~\ref{fig:predictive_test_full} in Appendix~\ref{app:predictive_coverage_test}). Overall, the network shows good performance when recovering the ground truth values of the parameters. However, we observe that it struggles to infer $\mu_{\log B, 2}$, $\sigma_{\log B, 2}$ and $w_{\log B}$ which characterize the second log-normal component of the initial magnetic-field distribution. This may be due to the limited size of the X-ray sample, whose low-number statistics introduce substantial statistical fluctuations in the input maps.
Moreover, we compute the coverage probability over the same test dataset (see Fig.~\ref{fig:coverage_plots} in Appendix~\ref{app:predictive_coverage_test}), which shows that the approximated posteriors are conservative since the coverage probability is overall greater than the credibility level \citep[][see Appendix~\ref{app:predictive_coverage_test} for more details]{Hermans2021, Graber2024, Pardo-Araujo2025}. This further highlights the robustness of the inference results. 

To visualize the best-fit model, we sample 100 parameter sets from the posterior distribution and use our population synthesis framework to generate the respective simulations. We then compare the results of these simulated populations with the real observed populations. In Fig.~\ref{fig:PPdot_sim_vs_obs}, we show the comparison between our best simulations and the observed dataset in the $P-\dot{P}$ diagram. The contour lines represent the distribution of the simulated populations obtained via Kernel density estimation (KDE), plotted over the real observed radio pulsars (dots) and X-ray emitting neutron stars (crosses). Both the radio pulsar population and the entire X-ray emitting neutron star population are reasonably well reproduced by the simulations.

However, if we focus our attention on the sub-populations of young magnetars and XDINSs, our simulations struggle to reproduce the observations. To define the sub-population of young magnetars, we follow \cite{Pardo-Araujo2026} and select those sources from the observed dataset whose ages (estimated from a supernova remnant association or from the characteristic age) are lower than 2 kyr (pink crosses). To compare with this sub-sample, in our simulations we then select the neutron stars with simulated ages lower than 2 kyr and fit a KDE (pink contour lines). For the sub-population of XDINS-like sources (orange contour lines), we consider the simulated neutron stars within a distance of $\sim 500$ pc from the Sun and an age greater than $10^5$ yr to mimic the properties of the observed XDINSs (orange crosses). We then compute the mean and standard deviation of the number of detected sources in these two sub-populations over 100 simulations. While the simulations are able to predict a number of detected young magnetars, i.e., $6 \pm 3$, which is compatible with the seven observed sources, our simulation framework predicts their distribution to be shifted towards lower spin periods and spin-period derivatives. On the other hand, for the XDINSs, the simulations are able to reproduce the observed timing properties but our model underestimates their number to be $2 \pm 2$ compared to the seven observed sources.

By counting the total number of neutron stars we need to generate per simulation to reach the observed count in each survey (see Section~\ref{sec:observed_sample}), we can estimate a neutron star birth rate for each of the simulated surveys. The estimated mean and standard deviation are as follows:
\begin{align} \label{eqn:BR_estimated}
&\text{\ac{PMPS}: $\sim 3.6 \pm 0.8$ neutron stars per century}, \nonumber \\
&\text{\ac{SMPS}: $\sim 3.0 \pm 0.7$ neutron stars per century},  \\
&\text{\ac{HTRU}: $\sim 3.0 \pm 0.7$ neutron stars per century}, \nonumber \\
&\text{X-ray: $\sim 3.1 \pm 0.9$ neutron stars per century}. \nonumber
\end{align}
Note that these values are compatible with each other at the 1-$\sigma$ level but are higher than the core-collapse supernovae rate of $1.63 \pm 0.46$ per century estimated by \cite{Rozwadowska2021}. 

In Fig.~\ref{fig:logN-logS_sim_vs_obs}, we compare the observed $\log N -\log S$ distribution for the absorbed X-ray fluxes (black line) with that predicted by the simulations, where the purple line denotes the median prediction and the shaded bands indicate the 1-$\sigma$ and 3-$\sigma$ uncertainties. In this plot, $N$ represents the number of sources with an absorbed X-ray flux larger than a given value $S_{\rm X,abs}$. The observed $\log N -\log S$ is marginally consistent with our simulations, falling inside the 3-$\sigma$ uncertainty band. However, in our simulations we observe a lack of bright sources with fluxes above $\sim \unit[10^{-12}]{erg \, s^{-1} \, cm^{-2}}$. Since the tail at high fluxes is dominated by young and/or nearby bright sources, a regime where the completeness level of the observed sample should be high, this lack of sources in our simulations is another indication that the model either underestimates the detected numbers or the observed fluxes for the two sub-populations summarized above (see Section~\ref{sec:discussion} for further discussion of this aspect).

\subsection{Inference on the sub-sample of young magnetars and XDINSs}
\label{sec:results_youngxdins}


\begin{figure*}
\centering
\includegraphics[width = 1\textwidth]{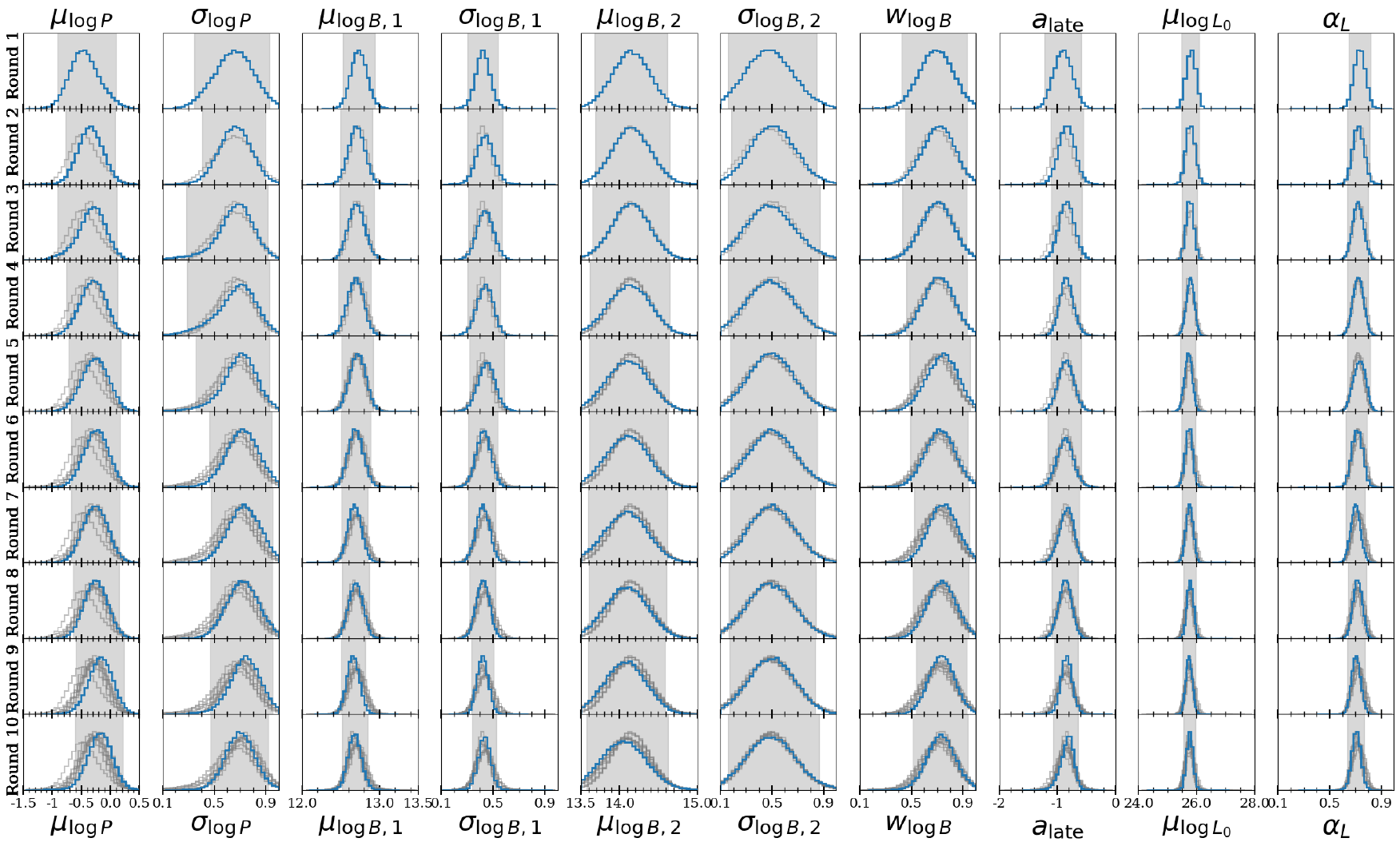}
\caption{Results of the \ac{TSNPE} algorithm applied to our population synthesis considering the sub-samples of young magnetars and XDINSs (see Section~\ref{sec:results_youngxdins}). Each row shows the marginalized posteriors obtained in each round of inference where the columns refer to different parameters describing the log-normal birth spin-period distribution ($\mu_{\log P}$, $\sigma_{\log P}$), the double log-normal birth magnetic-field distribution ($\mu_{\log B,1}$, $\sigma_{\log B,1}$, $\mu_{\log B,2}$, $\sigma_{\log B,2}$, $w_{\log B}$), the power-law index for the late time decay of the magnetic field ($a_{\rm late}$), and the radio luminosity prescription ($\mu_{\log L_0}$, $\alpha_L$). For a given parameter, the marginalized posterior computed in that specific round is shown in blue, while the marginalized posteriors from previous rounds are shown in light gray. The gray shaded area represents the $95\%$ credibility interval of the approximated posterior for that round. In each of these 1D marginal posterior distribution, the horizontal axes represent the parameters’ prior ranges.}
\label{fig:marginal_posterior_B_double_lognorm_dip-tor_heavy_youngxdins}
\end{figure*}



\begin{figure*}
\centering
\includegraphics[width = \textwidth]{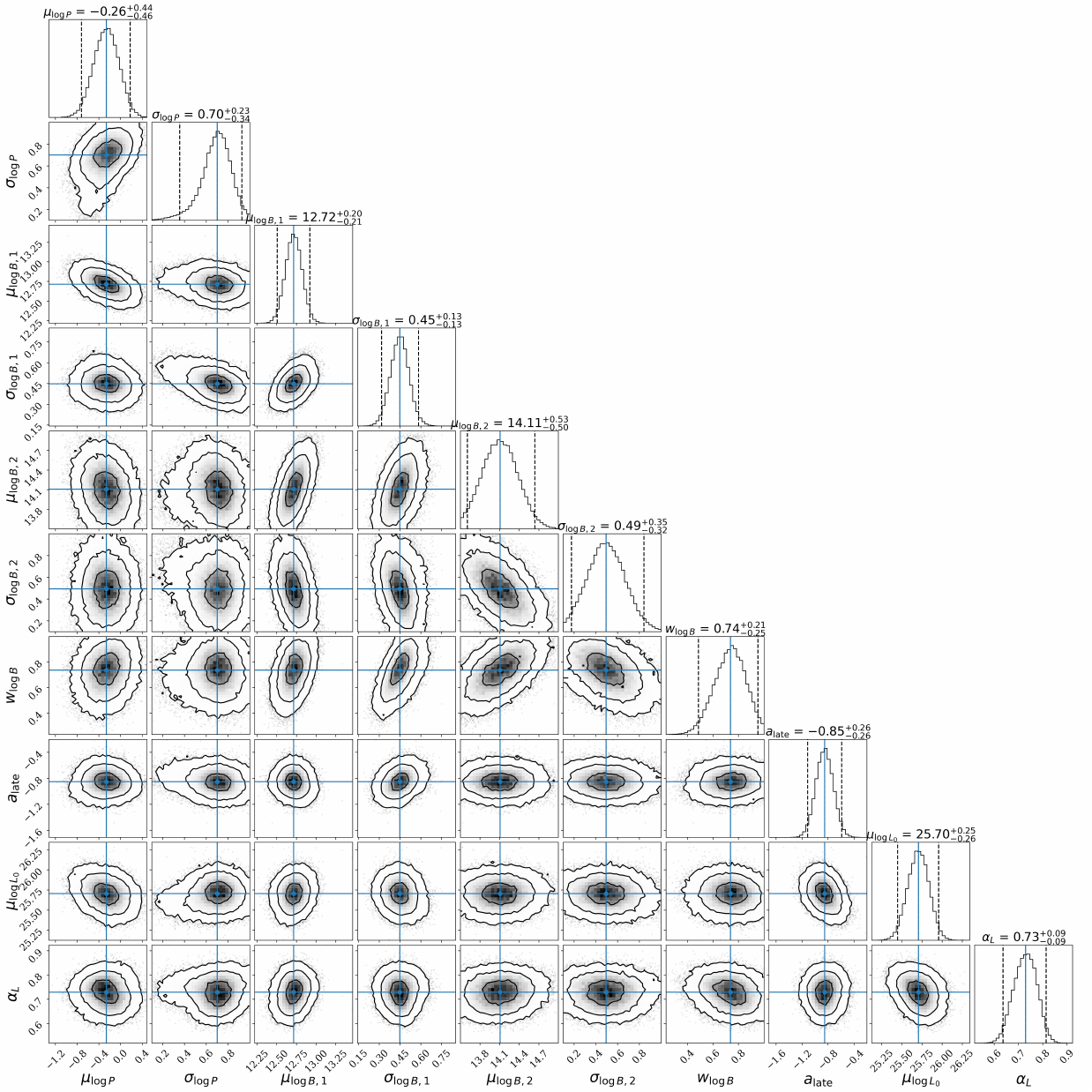}
\caption{Corner plot of the posterior distribution conditioned on the observed data restricted to the radio pulsars and the sub-populations of young magnetars and XDINSs (see Section~\ref{sec:results_youngxdins}). The posterior is estimated from round 5 of the TSNPE algorithm and adopted for the best-parameter analysis in Section~\ref{sec:results_youngxdins}. The parameters describe the log-normal birth spin-period distribution ($\mu_{\log P}$, $\sigma_{\log P}$), the double log-normal birth magnetic-field distribution ($\mu_{\log B,1}$, $\sigma_{\log B,1}$, $\mu_{\log B,2}$, $\sigma_{\log B,2}$, $w_{\log B}$), the power-law index for the late time decay of the magnetic field ($a_{\rm late}$), and the radio luminosity prescription ($\mu_{\log L_0}$, $\alpha_L$). The contour levels in the 2D marginalized posteriors represent the 1-$\sigma$, 2-$\sigma$ and 3-$\sigma$ credible regions. For each parameter, we report the median value and the 2-$\sigma$ credible interval. In the 1D marginalized posterior distributions, these are marked by the blue solid line and the dashed lines, respectively.}
\label{fig:corner_plot_youngxdins}
\end{figure*}



\begin{figure*}
\centering
\includegraphics[width = 0.5\textwidth]{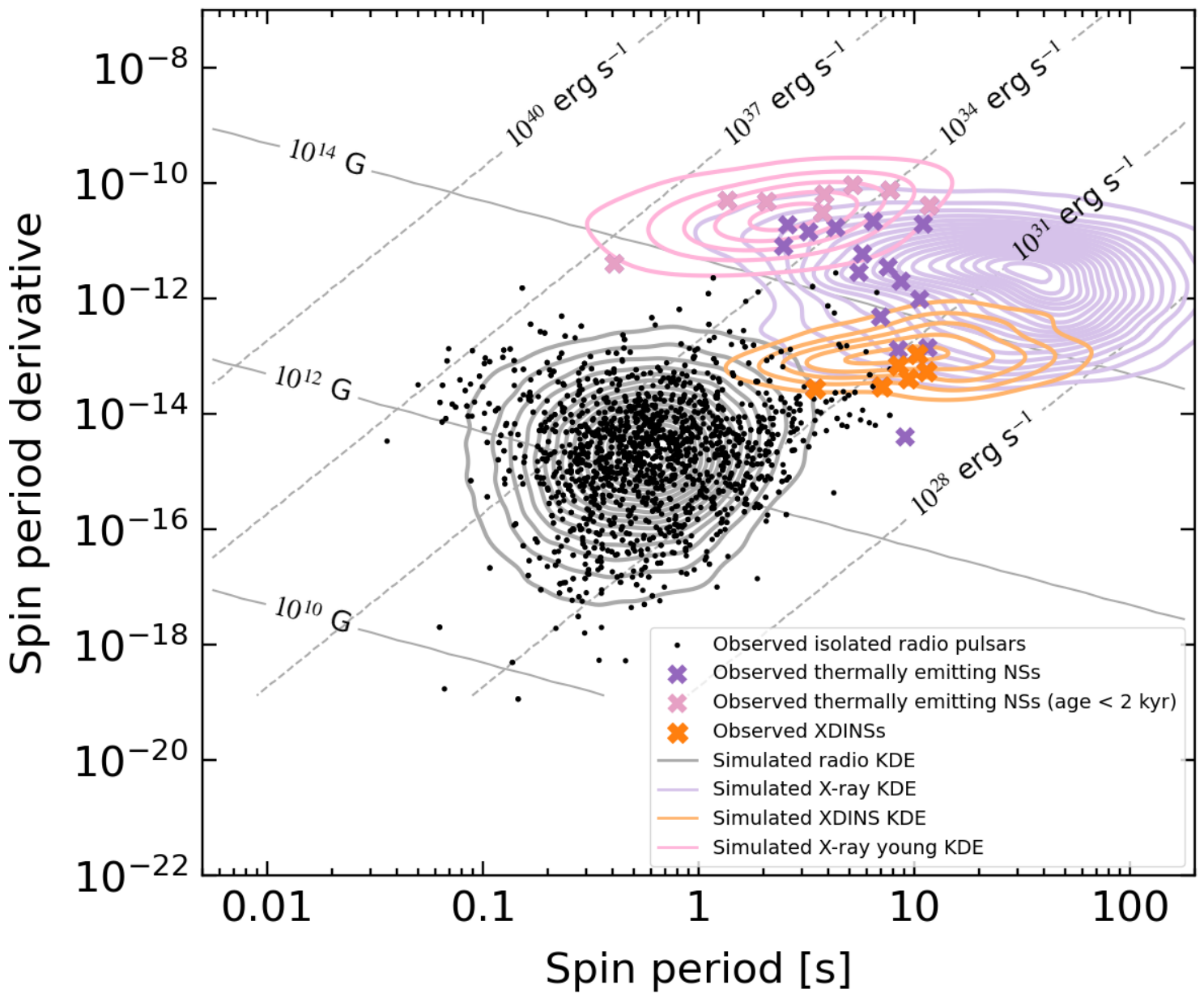}
\caption{Comparison in the $P-\dot{P}$ plane between the observed population of radio pulsars and X-ray emitting neutron stars and the best simulated populations for the experiment where we consider the sub-sample of young magnetars and XDINSs (see Section~\ref{sec:results_youngxdins}). Points and crosses represent the observed radio pulsars and X-ray emitting neutron stars, respectively. Among the X-ray-emitting neutron stars, pink crosses indicate young magnetars (estimated ages $<2$ kyr), while orange crosses denote the observed XDINSs. The contour lines show KDE density contours derived from 100 simulations with input parameters sampled from the round-5 posterior distribution. The pink and orange contours correspond to the two simulated sub-populations, while the purple contours represent the full simulated X-ray-emitting neutron star population.}
\label{fig:PPdot_sim_vs_obs_youngxdins}
\end{figure*}



\begin{figure*}
\centering
\includegraphics[width = \textwidth]{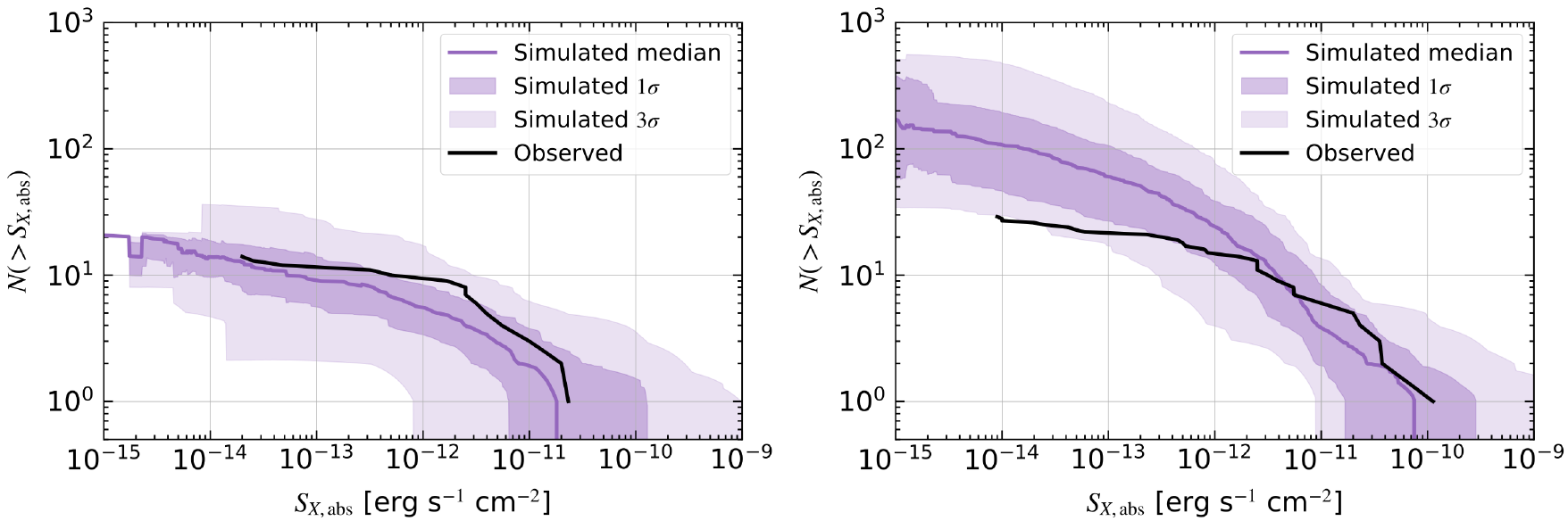}
\caption{Comparison of the $\log N-\log S$ distributions for the observed and simulated thermally emitting X-ray neutron stars for the experiment where we consider the sub-sample of young magnetars and XDINSs (see Section~\ref{sec:results_youngxdins}). The left panel considers only the young magnetars (with ages lower than 2 kyr) and XDINS-like sources. The right panel represents the entire observed X-ray population. In both plots, the black lines represent the trend for the observed population, while the purple lines and bands represent the median, the 1-$\sigma$ and 3-$\sigma$ uncertainties, respectively. The latter are computed over 100 simulations generated from sets of parameters drawn from the posterior distribution obtained from round 5 of our TSNPE algorithm.}
\label{fig:logN-logS_sim_vs_obs_youngxdins}
\end{figure*}



\begin{figure}
\centering
\includegraphics[width = 0.45\textwidth]{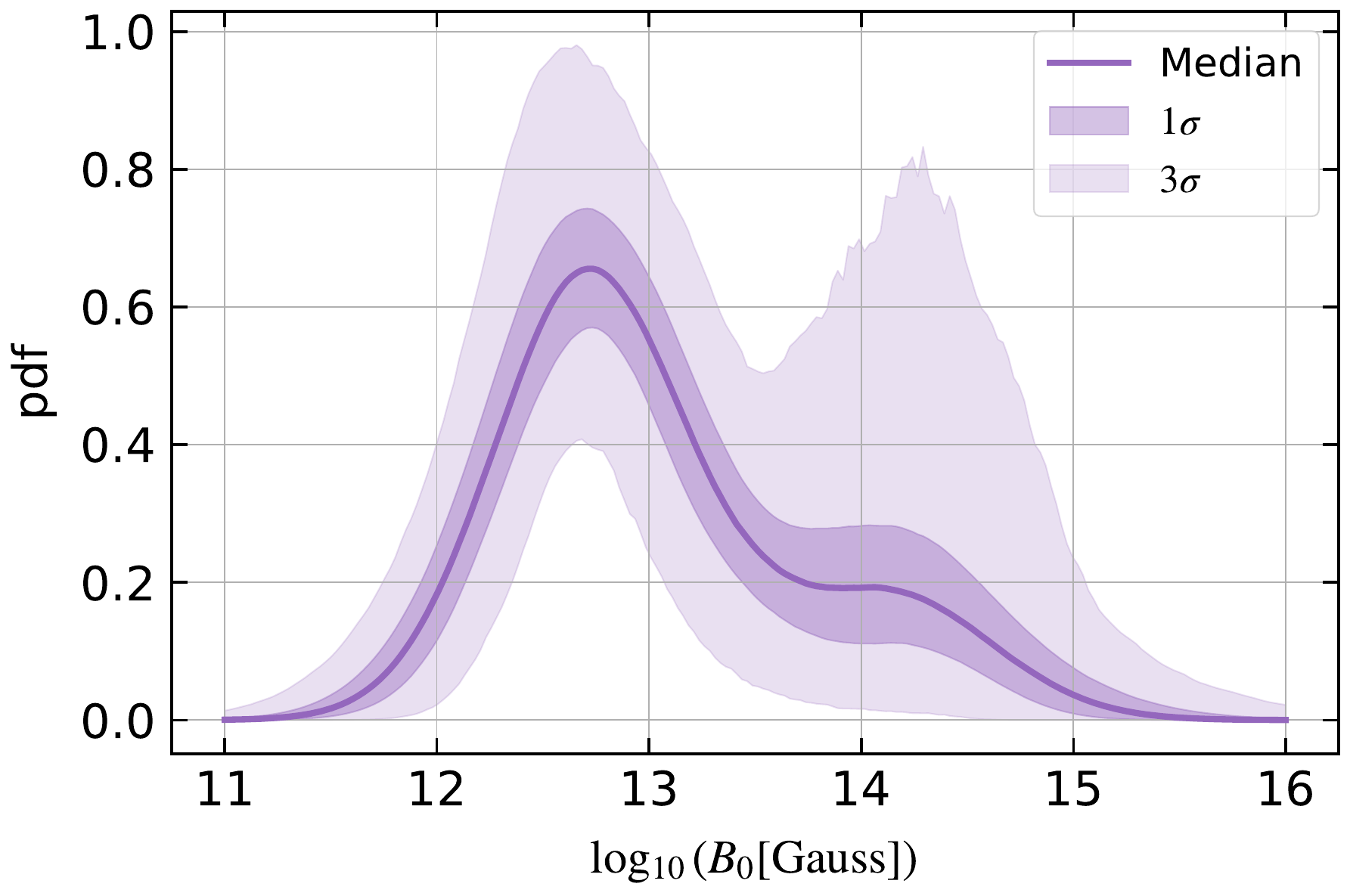}
\caption{Best fit distribution of initial magnetic fields for the inference experiment on radio pulsars, young magnetars and XDINSs (see Section~\ref{sec:results_youngxdins}). The purple solid line represents the median value, while the shaded regions represent the $1$-$\sigma$ and $3$-$\sigma$ uncertainties computed over 10000 random samples of the posterior distribution shown in Fig.~\ref{fig:corner_plot_youngxdins}.}
\label{fig:initialB_youngxdins}
\end{figure}



\begin{figure}
\centering
\includegraphics[width = 0.45\textwidth]{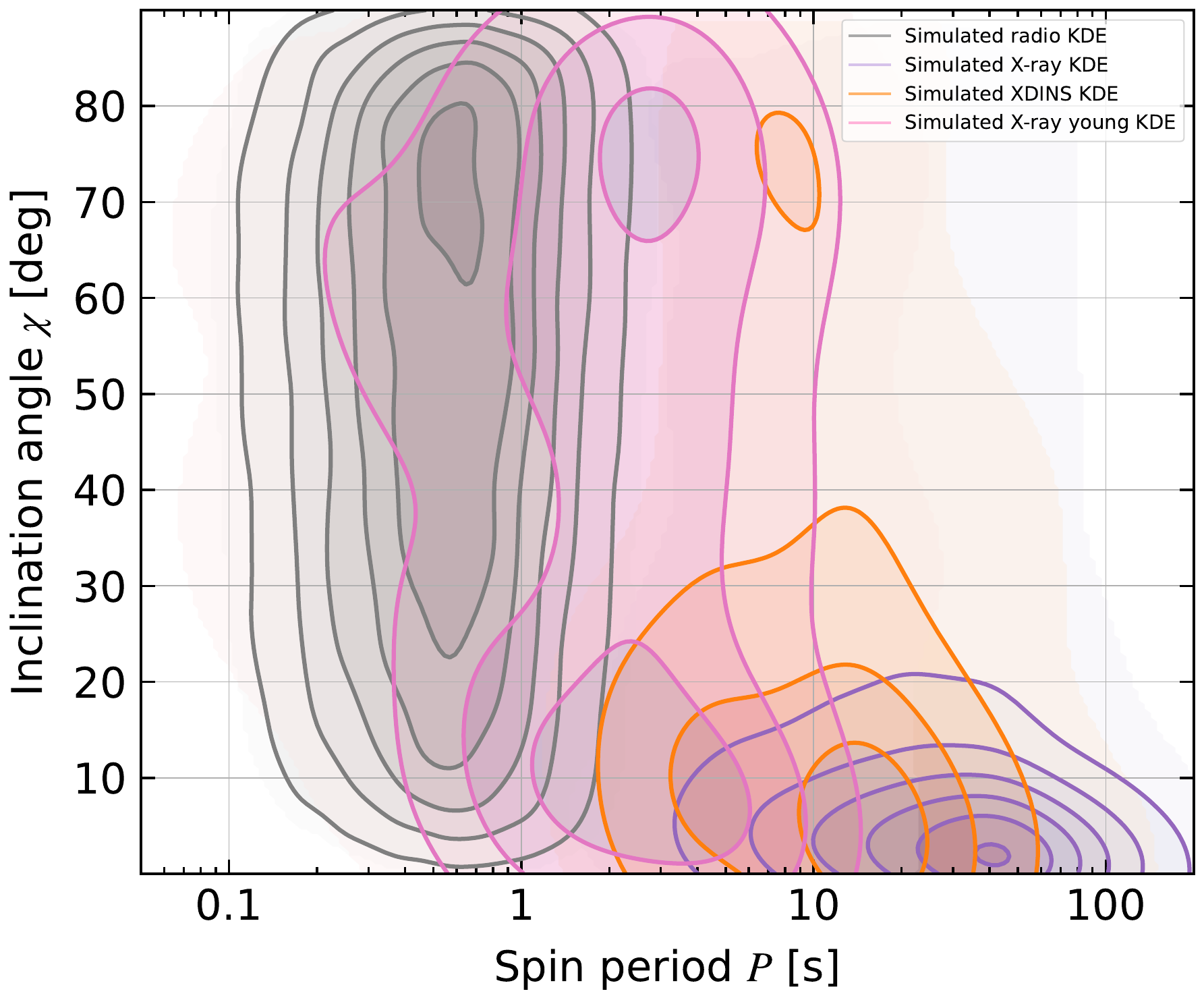}
\caption{Distribution of the final magnetic inclination angle, $\chi$, as a function of the final spin period for the simulated best-parameter populations for the experiment where we consider the sub-sample of young magnetars and XDINSs (see Section~\ref{sec:results_youngxdins}). The contour lines represent KDE density contours derived from 100 simulations using input parameters sampled from the round-5 posterior distribution. Gray and purple contours represent the simulated radio pulsars and X-ray emitting neutron stars, respectively. Pink contours represent the young (ages lower than 2 kyrs) magnetars, while orange contours denote the XDINS-like sources.}
\label{fig:inclination_angle}
\end{figure}


Motivated by the results presented above, we perform a second experiment and check the impact of considering only the restricted samples of young magnetars and XDINSs (as defined in the previous section) in addition to the radio pulsar sample on our inference results. These two sub-samples should be less biased and more complete. On one hand, young sources are X-ray bright, so that observational biases are not expected to significantly affect their detection. XDINSs, on the other hand, are observed to have high X-ray fluxes due to their proximity, representing an almost complete sample when considering a limiting distance of $\sim 500$ pc from the Sun \citep[][discuss possible new candidates]{Kurpas2026a}. We note that a new XDINS with an upper limit distance of 700 pc was recently confirmed \citep{Kurpas2026b}. As this result became available after the analysis presented here had been completed, this new source is not included in our observational sample. However, we note that its location in the $P-\dot{P}$ diagram is similar to the other XDINSs so that our results should not be affected. We thus account for a total of seven young magnetars and six XDINSs, leading to 13 X-ray bright sources available for inference in addition to the radio pulsar population. We generate the maps using these extra filters on the X-ray population for both the simulations and the observed sample and train the TSNPE algorithm. The inferred posterior distribution conditioned on these restricted observed data for each round are displayed in Fig.~\ref{fig:marginal_posterior_B_double_lognorm_dip-tor_heavy_youngxdins}. As in the previous experiment, the posterior distribution is overall stable across the training rounds. 

In order to compare the population simulated from the inferred posterior with the real one, we again use the posterior estimate from round 5 as it displays conservative results encompassing the ranges of posteriors estimated in the following rounds. In Fig.~\ref{fig:corner_plot_youngxdins}, we show the corner plot of the posterior distribution inferred in round 5, while in the second column of Table~\ref{tab:infer_results}, we list the corresponding medians of the inferred parameters together with the 95\% credible intervals. In this case, we find for the initial field distribution that the first log-normal component still peaks at $5\times 10^{12}$\,G, while the second one has moved to larger values, i.e., $1\times 10^{14}$\,G, compared to the previous experiment.
To assess the quality of the trained posterior estimator in this round, we have checked its predictive power over the test dataset and computed the coverage probability (see Appendix~\ref{app:predictive_coverage_test} and Figs.~\ref{fig:predictive_test_youngxdins} and \ref{fig:coverage_plots}). The network performance is overall similar to the previous experiment, again highlighting the robustness of the inference results. 

As in the previous section, we sample 100 parameter sets from the round-5 posterior distribution and generate the respective simulations. We then compare the distributions of these simulated populations with the real observed populations in the $P-\dot{P}$ plane (see Fig.~\ref{fig:PPdot_sim_vs_obs_youngxdins}). The radio pulsar population is again well recovered by the simulations. However, when looking at the X-ray population, while this time the young magnetars (pink crosses) and XDINSs (orange crosses) are well reproduced by the simulations, the bulk of the observed mock X-ray emitting neutron stars (purple contour lines) is shifted towards longer periods compared to real data.

The estimated mean and standard deviation of the birth rates for the different surveys are the following:
\begin{align} \label{eqn:BR_estimated_youngxdins}
&\text{\ac{PMPS}: $\sim 3.6 \pm 0.9$ neutron stars per century}, \nonumber \\
&\text{\ac{SMPS}: $\sim 3.0 \pm 0.8$ neutron stars per century},  \\
&\text{\ac{HTRU}: $\sim 3.0 \pm 0.7$ neutron stars per century}, \nonumber \\
&\text{X-ray: $\sim 1.3 \pm 0.9$ neutron stars per century}. \nonumber
\end{align}
We again find that the birth rate inferred for the radio surveys is larger than the predicted core-collapse supernovae rate from \cite{Rozwadowska2021}. However, while the values of the three radio surveys are still compatible with each other at the 1-$\sigma$ level, the mean birth rate for the X-ray survey is significantly lower compared to the others. This is an indication that this time, our simulations are overestimating the number of detected X-ray emitting neutron stars compared to reality.
However, while overestimating the numbers for the entire X-ray detected population, these simulations predict $8 \pm 5$ young magnetars and $4 \pm 3$ XDINS-like sources allowing a better agreement with the real observed numbers in these two sub-categories, i.e., seven and seven, respectively.

In Fig.~\ref{fig:logN-logS_sim_vs_obs_youngxdins}, we compare the $\log N -\log S$ distributions. In the left panel, we show the distribution when considering only the young magnetars and XDINS-like sources. In this case, the observed $\log N -\log S$ falls almost entirely inside the 1-$\sigma$ uncertainty band of the simulations. When considering the whole X-ray population (right panel), the simulations are compatible at the 1-$\sigma$ level with the high-flux tail of the observed distribution. However at low fluxes, while still being compatible with the observed distribution at a 3-$\sigma$ level, the simulations predict more sources at flux values between $\unit[10^{-14} - 10^{-12}]{erg \, s^{-1} \, cm^{-2}}$. This excess is also reflected in the $P-\dot{P}$ diagram where the number of detected mock sources with such low fluxes and spin periods in the range $10-100$\,s is overestimated, as highlighted above. Nevertheless, it is important to note that the simulations are able to reproduce the tail at high fluxes where young or nearby bright sources dominate and where the completeness of the observed sample is high.

In Fig.~\ref{fig:initialB_youngxdins}, we show the best-fit distribution for the initial magnetic field together with its uncertainty. In particular, we draw 10000 values from the posterior distribution obtained from round 5 (see Fig.~\ref{fig:corner_plot_youngxdins}) and evaluate the median distribution (purple solid lines) and the $1-\sigma$ and $3-\sigma$ uncertainties (shaded regions).
We note that under our assumptions, reproducing the properties of both the young magnetar and XDINS populations requires a birth magnetic field distribution with a second component peaking around $10^{14}$ G and extending up to $10^{15}$ G. Given the uncertainties on the weight parameter, $w_{\log B}$, this second component may account for as much as 50\% of the neutron-star population at birth.

Finally in Fig.~\ref{fig:inclination_angle}, we show the distribution of final inclination angles between the magnetic and rotation axes as a function of final spin periods for the different sub-populations obtained using the same best-parameter simulations as shown in Fig.~\ref{fig:PPdot_sim_vs_obs_youngxdins}. Contour lines and color shading represent density levels obtained via KDE. Owing to their weak magnetic fields and negligible magnetic-field decay, radio pulsars (gray) undergo slow and rather inefficient alignment (see Eq.~\eqref{eq:chidot}). 
Moreover, the detected radio pulsars are biased towards larger inclination angles because such configurations are more likely to beam towards the observer. Consequently, the inclination angles of these sources remain broadly distributed, with a median and 1-$\sigma$ percentiles of $51_{-29}^{+25}$\,deg. In contrast, the detected X-ray population (purple) exhibits a stronger tendency towards alignment because the high magnetic fields lead to shorter alignment timescales. In addition, X-ray detection is not affected by the same beaming bias as noted for the radio population, resulting in smaller $\chi$ values. While similarly broad, the distribution of inclination angles for the X-ray population exhibits a median and 1-$\sigma$ percentiles of $6_{-5}^{+45}$\,deg. Although young magnetars (pink) possess the strongest magnetic fields, they have not had sufficient time to evolve towards alignment, resulting in a broad inclination angle distribution with a median and 1-$\sigma$ percentiles of $42_{-35}^{+34}$ deg. Finally, XDINS-like sources (orange), being typically older, are instead more aligned, with a median and 1-$\sigma$ percentiles of $12_{-12}^{+45}$ deg.


\section{Discussion}
\label{sec:discussion}

We first turn our attention to the spin-period parameters. Compared with the radio-pulsar only results obtained in \cite{Pardo-Araujo2025}, here we find that the initial spin period distribution is shifted to somewhat larger values, although the values are consistent within the 95\% credible interval. The corresponding uncertainties on the spin-period parameters are still large and indicate that these parameters are difficult to constrain. As already discussed in previous works, this is due to the fact that the memory of the initial spin period is rapidly lost during magneto-rotational evolution. Adding the sample of X-ray emitting neutron stars with higher magnetic fields does not help in this regard, because the stronger the magnetic field, the faster the evolution in the $P-\dot{P}$ diagram.

As $\mu_{\log P}$ is negatively correlated with the initial magnetic field parameters, $\mu_{\log B,1}$ and $\mu_{\log B,2}$ (see Fig.~\ref{fig:corner_plot_youngxdins}), an initial spin period distribution shifted to larger values allows for lower values of $\mu_{\log B,1}$.
Hence, the first Gaussian component of the magnetic field distribution is shifted to slightly lower values compared to \cite{Graber2024, Pardo-Araujo2025}. This effect is also mitigated by the addition of the second component at stronger magnetic fields that provides detectable radio pulsars with magnetic fields above $10^{13}$\,G. 

Furthermore as in \cite{Pardo-Araujo2025}, we find a slightly negative correlation between the radio luminosity parameter, $\mu_{\log L_0}$, and $\mu_{\log P}$, and a slightly positive correlation between $\mu_{\log L_0}$ and $\mu_{\log B,1}$. This means that as the radio pulsar population shifts to the right and bottom of the $P-\dot{P}$ plane, the radio luminosity normalization factor takes on smaller values compared with previous studies because of our larger $\mu_{\log P}$ and smaller $\mu_{\log B,1}$. Otherwise too many pulsars with low rotational energies would be detected.

It is interesting to note that the inferred value for the $a_{\rm late}$ parameter is consistent within uncertainties with that obtained in \cite{Pardo-Araujo2025}. This parameter mainly regulates the evolution of the $\dot{P}$ values at times around $10^6$ yr and later. $a_{\rm late}$, therefore, predominantly affects the old radio pulsars and XDINS population. Obtaining a consistent value across different neutron star classes indicates that the magnetic field evolution model we adopt is able to reproduce the properties of both classes simultaneously.

Turning to the parameters relevant for the X-ray population, we find that when inferring on the entire sample of X-ray emitting neutron stars, the second component of the initial magnetic field is centered at $\sim \unit[7 \times 10^{13}]{G}$.
On the other hand, when focusing the inference on the sub-populations of young magnetars and XDINSs, the mean value of this component shifts to $\sim 1 \times 10^{14}$ G.
This difference is due to the fact that in the first case, when fitting the entire observed X-ray population, the TSNPE algorithm tries to reproduce the shape and location of the bulk of the population in the $P-\dot{P}$ plane which shows a clear cut off at periods of around $\sim \unit[10-20]{s}$ (see Fig.~\ref{fig:PPdot_sim_vs_obs}). To achieve this, the initial magnetic field cannot be too large. Otherwise, the spin-down would be too strong, leading to final spin periods larger than those observed. However, by limiting the initial magnetic field strength this way, the simulations poorly reproduce the timing properties of the observed young magnetars as can be seen from the pink contour lines in Fig.~\ref{fig:PPdot_sim_vs_obs}, which are shifted to smaller values for both $P$ and $\dot{P}$ compared to the observed data (pink crosses). Another effect of restricting the mean magnetic field of the second component is the reduction of X-ray luminosities and observed fluxes, which is evident in the high-flux tail of the $\log N - \log S$ distribution (see Fig.~\ref{fig:logN-logS_sim_vs_obs}). 
On the other hand, when focusing the inference on the young magnetars and XDINSs, a second component with stronger initial magnetic fields can reproduce both sub-populations. This also implies higher thermal luminosities and a better fit to the high-flux tail of the $\log N - \log S$ distribution (see Fig.~\ref{fig:logN-logS_sim_vs_obs_youngxdins}). However, the resulting bulk population of X-ray emitting neutron stars tends to have spin periods longer than those observed due to a stronger spin-down (see purple contour lines in Fig.~\ref{fig:PPdot_sim_vs_obs_youngxdins}). 

In general, the uncertainty on the parameters describing the second component of the initial magnetic field distribution is large as can be observed by the component's significant variance in Fig.~\ref{fig:initialB_youngxdins}. The network struggles to predict the parameters $\mu_{\log B,2}$, $\sigma_{\log B,2}$ and the weight parameter, $w_{\log B}$, as can be seen in Figs.~\ref{fig:predictive_test_full} and ~\ref{fig:predictive_test_youngxdins} in Appendix~\ref{app:predictive_coverage_test}. This is a consequence of the low-number statistics for the X-ray neutron star sample which can introduce random statistical noise in the maps of the simulated and observed sources making the training process more challenging and the inference results less precise.
In addition, this uncertainty could also be an indication that the functional form describing the initial magnetic field distribution, the adopted models for the evolution of magnetar-like sources in the $P-\dot{P}$ diagram or the observational biases are not fully capturing reality. 
In particular, we note that \cite{Shimasue2026} recently proposed that the observed dichotomy between radio pulsars and magnetars may be driven by observational selection effects alone. Specifically, stronger alignment of the magnetic and rotation axes in the magnetar population can reduce the beaming fraction and, hence, the likelihood of detecting these strong $B$-field sources in the radio band, an effect also incorporated in our simulations (see also Section~\ref{sec:lpt}). By correcting for these effects, \citeauthor{Shimasue2026} argue that a single continuous initial birth magnetic-field distribution like a log-uniform may explain both populations better. Performing Bayesian model comparison studies and enlarging the sample of detected high-$B$ pulsars in the gap between radio pulsars and magnetars would be crucial for shedding more light on these aspects.

We note that our simulations predict a number of XDINS-like sources which is only marginally consistent with observations and slightly underestimate it. There are several reasons for this discrepancy. 
First of all, due to the proximity of these neutron stars, a number estimate is highly dependent on the precise modeling of the solar neighborhood for which we only consider a simplified, averaged description (see Section~\ref{subsec:dynamical_evolution}). However, the Sun is located in a region that has experienced enhanced recent massive-star formation, historically associated with the Gould Belt \citep{Alves2020, Zucker2022}. This implies a local core-collapse supernova and thus neutron-star birth rate, which might exceed the Galactic average. A more detailed modeling of the massive-star distribution in the vicinity of the Sun may hence help refine our predictions of XDINS-like sources \citep[see, e.g.,][]{Posselt2008}. 
A second aspect to consider is that we do not account for the dependence of magneto-thermal evolution on the neutron-star mass. In particular, for some equations of state (including BSk24, which we adopt here), sufficiently massive neutron stars may exceed the threshold for the onset of direct Urca neutrino emission, leading to enhanced cooling and consequently reduced detectability \citep[see, e.g.,][]{Marino2024}. This effect could generally alter the predicted detection numbers and the X-ray flux distribution. However, quantifying the fraction of neutron stars undergoing this evolutionary channel is challenging due to the large uncertainties on the neutron star mass distribution \citep[see, e.g.,][]{You2025, Basu2025}.

Finally, it is worth mentioning that our simulation framework is agnostic to the formation environment of isolated neutron stars and does not model binary and isolated formation channels separately. As most neutron star progenitors are born in binary systems \citep{Sana2012}, it is reasonable to expect that the birth properties (e.g., birth magnetic fields, birth spin periods, kick velocities and masses) of a fraction of presently isolated neutron stars has been shaped by the evolutionary history of the binaries from which they originated \citep[see, e.g.,][]{Hu2026}. While quantifying the impact of the binary formation channel on the observed population is outside the scope of this work, it may represent an additional source of uncertainty in the predicted numbers and properties, particularly for the detected young magnetars and XDINSs, owing to their small-number statistics.

\subsection{The $\sim 20$ s cut-off in the magnetar population}

The presence of a cut-off in the observed spin period distribution at around $\sim \unit[20]{s}$ is intriguing and has been interpreted as evidence for the presence of a highly resistive layer within the crust \citep{Pons2013}, generally associated with the nuclear pasta region.
However, this cut-off has been challenging to explain in population synthesis studies \citep[see for example][]{Gullon2015}. The fact that our framework also struggles to reproduce this sharp feature is due to different aspects. 

First of all, our treatment of the outburst rate is simplified, primarily as we rely on 2D magneto-thermal simulations where failures are obtained under simplified assumptions \citep[see][for more details]{Dehman2020}. Secondly, we assume that all crust failure events lead to a detectable outburst. This picture does not take into account the energy of these events and their luminosity in gamma-rays and X-rays. Quantifying the efficiency with which the energy dissipated by the stresses is converted into photon luminosity that could be detected by gamma-ray and X-ray instruments is challenging \citep{Lander2023, Qu2026}. Furthermore, we are not taking into account sensitivities and limitations of current and past X-ray observatories for transient detection. One issue, for example, is that outbursts occurring during periods when the source is Sun-constrained, i.e., too close to the Sun to be observed, will be missed. This happens for a few months each year for every source, effectively reducing the detectability time window. Therefore, accounting for these aspects, our simulations are likely overestimating the number of sources that could be detectable following an outburst event.

A second aspect to consider is that our results rely on the assumption that the configuration of the initial magnetic field has only dipolar poloidal and toroidal components. For the magnetar population, however, this is an oversimplification as the presence of strong multipole components in addition to the dipole \citep[e.g.][]{Dehman2023b} affects how the magnetic field evolves. This may lead to a faster decay and an enhanced dissipation in the neutron star crust that could lead to higher X-ray luminosities in the first stages of magnetar evolution. A faster decay could help explain the sharp drop of sources at a spin period of $\sim 20$\,s because the evolutionary spin-down trajectory would bend down sooner in the $P-\dot{P}$ diagram. However, we also note that a different configuration of the field would also affect the rate of expected crustal failures and consequently of outburst events. The presence of strong multipolar components in general should enhance the stresses in the crust and the rate of failures leading to more outburst events and a greater number of detectable magnetars.

Another cut-off affecting assumption is the evolutionary model. In the case of young magnetars, their frequent flares and outburst activity could cause changes in the magnetospheric configuration and drive particle winds that could momentarily enhance the spin-down rate \citep{Tong2013, Petri2019}. In this case, assuming a force-free dipolar magnetospheric spin-down model could be an over-simplification and bias the estimate of the surface dipolar magnetic field of young magnetars, and therefore, their birth properties. A different evolutionary track in the $P-\dot{P}$ diagram, where $\dot{P}$ is enhanced could also contribute to explaining the timing properties of magnetars. However, an enhanced spin-down acting over a long time would lead to longer final spin periods, which would be against the observed cut-off at $\sim 20$\,s. On the contrary, if the enhanced spin-down is only temporary, this would imply that we are detecting magnetars only in this short phase, which would make such detections more unlikely unless strong detection biases, like an outburst event, are present. 

An additional alternative evolutionary scenario is that fallback accretion shortly after the supernova modifies the stars' early spin and magnetic-field evolution. In this picture, the accreted material can temporarily bury the external dipolar magnetic field, reducing the spin-down torque until the field gradually re-emerges through diffusion on timescales of $10^3 - 10^5$\,yr \citep[e.g.,][]{Muslimov1995, Ho2011, Vigano2012, Bernal2013}. Such an evolutionary path has been proposed as a possible contributor to the diversity of isolated neutron-star populations \citep[e.g.,]{Fu2013, Popov2015, Torres-Forne2016, Rogers2016, Liu2019, Gourgouliatos2020} and could influence the spin-period distributions of magnetars and XDINSs. Quantifying its impact, however, would require coupling population synthesis with models of magnetic-field burial and re-emergence, which is beyond the scope of this work.

\subsection{The long-period pulsar sub-population}
\label{sec:lpt}


\begin{figure}
\centering
\includegraphics[width = 0.45\textwidth]{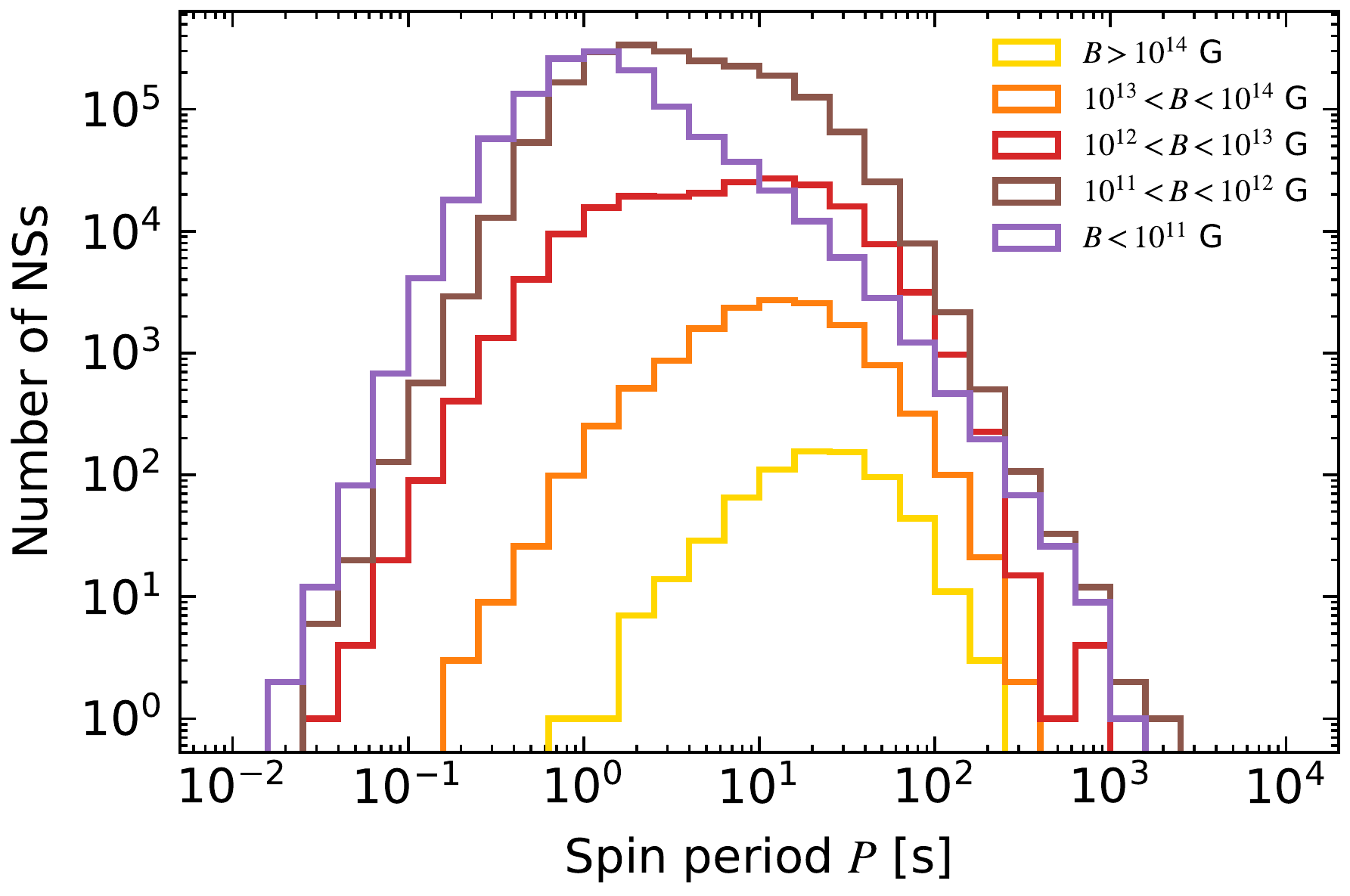}
\caption{Distribution of final spin periods for the entire evolved neutron star population for different ranges of final magnetic fields.}
\label{fig:LPT_prediction}
\end{figure}


Recently, a pulsar with a spin period of $\sim 76$ s \citep{Caleb2022} and several \acfp{LPT}, repeating on timescales ranging from hundreds of seconds up to several hours, have been discovered in radio \citep[see][for a review]{Rea2026}. Neutron stars, and in particular magnetars, have been suggested as possible engines to produce the emission of some of these LPTs. Based on our best-fit model discussed in Section~\ref{sec:results_youngxdins}, we can investigate this hypothesis and assess how many neutron stars in our Galaxy have ultra-long spin periods. To simulate the full population, we assume a birth rate of 3.5 neutron stars per century in line with Eq.~\eqref{eqn:BR_estimated_youngxdins} and evolve $3.5 \times 10^{6}$ neutron stars up to a maximum age of $10^8$\,yr. In Fig.~\ref{fig:LPT_prediction}, we produce histograms of the final spin periods considering different ranges of final magnetic fields for the entire neutron star population. We predict that the total number of sources (i.e., without taking into account any observational biases) with spin periods above 20\,s $\sim 2 \times 10^5$ ($\sim 6$\% of the entire population) of which $\sim 400$ objects ($\sim 0.01$\% of the entire population) have magnetar-like magnetic fields above $10^{14}$\,G. Moreover, we expect the number of neutron stars with spin periods above 100\,s to be $\sim 5000$ ($\sim 0.1$\% of the full population).

If we further consider the observational biases for the radio detection, we note that neutron stars with spin periods longer than a few seconds become hard to detect in our simulation framework. This is due to a combination of factors. First of all, we assume a radio beam aperture which has an inverse dependence on the spin period, i.e., $\propto P^{-1/2}$ \citep[see Eq.~(13) in][and references therein]{Graber2024}. Moreover, neutron stars reaching long spin periods are usually those with stronger magnetic fields and have therefore a tendency of evolving towards being aligned rotators, i.e., having magnetic dipole axes that are aligned with the rotation axes (see Eq.~\eqref{eq:magrot_evol}). This aspects naturally explain the dichotomy between radio pulsars and magnetars and why most magnetars are not observed as pulsating radio emitters \citep[see also][]{Shimasue2026}. Indeed, for slow rotators, the probability of the radio beam intersecting our line of sight (i.e., the beam fraction) is small, thus, reducing the possibility of detecting these sources as LPTs.
Furthermore, without considering alternative evolutionary spin-down models such as the interaction with supernova fallback disk \citep[see, e.g.,][]{Ronchi2022, Gencali2024, Zhou2024} or going beyond the standard dipolar spin-down picture, we cannot recover many sources that reach spin periods around 1000\,s. Therefore, without other mechanisms to enhance spin-down, our models can explain only the fast-spinning end of the LPT population.

\subsection{The source populations of FRBs}

There is broad and convincing evidence that FRBs are emitted by young, energetic neutron stars \citep[e.g.,][]{Bochenek2020, Pastor-Marazuela2025}. 
As the bursts are, on average, 10$^{12}$ times brighter than normal pulsars \citep[cf.][]{vanLeeuwen2023}, the inferred energy budget suggests that magnetars specifically could be emitting these bursts \citep[see, e.g.,][]{Zhang2020, ZhangJ2025, Shah2026}. 
This hypothesis is strengthened by evidence both at single-object and population level: the detection of a short bright burst from magnetar SGR~1935+2154 \citep{Bochenek2020}, the agreement in burst wait-time statistics between some repeating FRBs and magnetars \citep[e.g.,][]{Wadiasingh2019}, and the finding that the number density in cosmological volume, spectral index and dispersion measures of the population of FRBs agrees qualitatively with the expectations for a magnetar origin \citep{Gardenier2021}.
Our population synthesis method has the potential to make this link much more quantitative. 
By connecting our radio and X-ray results to the state-of-the-art in  
FRB population synthesis codes \citep[e.g.][]{Wang2024},
we may be able to determine, for example, if FRB repetition frequencies, too, are guided by the same outburst prescriptions that drive magnetar flares (see Section~\ref{sec:crust_failures}, and bottom panel of Fig.~\ref{fig:B_Lx_magneto-thermal}).  
Such a detection would offer strong evidence that FRBs are, indeed, distant magnetars.

\section{Conclusions}

We have performed a simulation-based inference study to constrain the Galactic population of neutron stars. Our approach is unique in constructing a unified framework for the populations of radio pulsars as well as magnetars and XDINSs which have a significant thermal component in their X-ray spectra. To achieve this, we develop \texttt{ML-Poppyns}, a population synthesis framework that combines models to simulate the birth properties, evolution and detection of a Galactic population of neutron stars, with simulation-based inference that exploits the power of neural networks to estimate model parameters. 
To reproduce the properties of magnetars and XDINSs, we incorporate results from state-of-the-art 2D magneto-thermal simulations to model the coupled evolution of magnetic fields, surface temperature and crustal failures. 

Using these ingredients, we then determine which model parameters best describe the observed populations.
Our results can help refine the expected detection numbers for future instruments like the SKA \citep{Keane2015, Keane2025} in the radio band and eXTP \citep{Santangelo2019} and NewAthena \citep{Cruise2025} in the X-ray band. Our framework can also serve as a starting point for investigating the impact of strongly magnetic neutron stars as central engines for powerful events such as FRBs and GRBs.

We show that an initial magnetic field distribution described by two log-normal components with means of $5 \times 10^{12}$ G and $1 \times 10^{14}$ G respectively, is able to reproduce the properties of radio pulsars, young magnetars and XDINSs. The best-fit model does, however, struggle to explain the absence of detected sources in the X-ray band with spin periods larger than $\sim 20$\,s. We also find that to simultaneously explain the detected numbers at both electromagnetic wavelengths, we require a Galactic neutron star birth rate of 3-4 neutron stars per century. 


\begin{acknowledgments}

We thank the anonymous referee for useful comments and suggestions that helped improve the manuscript.
M.R.\ and J.v.L. are supported by the Dutch Research Council (NWO) via the grant CORTEX (NWA.1160.18.316) of the research programme NWA-ORC. V.G.\ is supported by a UKRI Future Leaders Fellowship (grant number MR/Y018257/1). C.P.A.\ and N.R.\ are supported by the ERC via the Consolidator grant ``MAGNESIA'' (No.\ 817661), the ERC Proof of Concept ``DeepSpacePULSE'' (No.\ 101189496), and by the program Unidad de Excelencia Mar\'ia de Maeztu CEX2020-001058-M. We also acknowledge support from the Catalan grant SGR2021-01269 (PI: Graber/Rea) and the Spanish grant PID2023-153099NA-I00 (PI: Coti Zelati). C.P.A.’s work has been carried out within the framework of the doctoral program in Physics at the Universitat Autonoma de Barcelona. C.D. is supported by the Ministerio de Ciencia, Innovación y Universidades (JDC2023-052227-I), co-funded by AEI (MCIN/AEI/10.13039/501100011033), the FSE+, and the Universidad de Alicante. C.D. and J.P. acknowledge support from the Conselleria d'Educació, Cultura, Universitats i Ocupació de la Generalitat Valenciana (grant CIPROM/2022/13). C.D. acknowledges the allocation of computing resources provided by the Swedish National Allocations Committee at the Center for Parallel Computers at the Royal Institute of Technology in Stockholm (Sweden).
D.D.G.\ is supported by a Juan de la Cierva fellowship (JDC2023-052264-I). A.M. acknowledges support from the Fund Vera Rubin/Chile 2024, under the project DIA 1736 "Silent black holes around red supergiants". F.C.Z. is supported by a Ram\'on y Cajal fellowship (grant agreement RYC2021-030888-I).

The data production, processing, and analysis tools for this paper have been implemented and operated at the Port d’Informaci\'o Cient\'ifica (PIC) data center. PIC is maintained through a collaboration of the Institut de F\'isica d’Altes Energies (IFAE) and the Centro de Investigaciones Energ\'eticas, Medioambientales y Tecnol\'ogicas (Ciemat). We particularly thank Christian Neissner and Martin Børstad Eriksen for their support at PIC.

We made use of the pulsar population synthesis code ML-Poppyns \cite{Ronchi2021, Graber2024, Pardo-Araujo2025} funded by the European Research Council via the ERC Consolidator grant ``MAGNESIA'' (No.\ 817661; PI: N.\ Rea), and publicly available at \url{https://ice-csic-astroexotic.github.io/code/ml_poppyns/}.

The authors thank Roberto Turolla for useful exchanges on the X-ray emission and resonant cyclotron scattering and Emilie Parent for useful discussions on the radio detection.

\end{acknowledgments}


\vspace{5mm}

\software{Astropy \citep{astropy2013, astropy2018}, healpy \citep{Gorski2005, Zonca2019}, IPython \citep{PerezGranger2007}, JupyterLab, Matplotlib \citep{Hunter2007}, Numba \citep{Lam2015}, NumPy \citep{Oliphant2006, vanderWalt2011, Harris2020}, Pandas \citep{McKinney2010}, PyGEDM, PyTorch \citep{Paszke2019}, sbi \citep{Tejero-Cantero2020}, SciPy \citep{Jones2001, Virtanen2020}, Sphinx.}


\appendix

\section{Resonant cyclotron scattering}
\label{app:rcs}

The non-relativistic classical resonant scattering cross section for a photon with frequency $\omega$ can be approximated as \citep[see][]{Canuto1971, Nobili2008, Yamasaki2020}:
\begin{align} 
    \sigma_{\rm res}(\omega) \sim \pi^2 \frac{e^2}{m_{e} c} \delta(\omega - \omega_B) (1 + \cos^2 \theta),
\end{align}
where $e$ is the electron charge, $m_e$ is the electron mass, $c$ is the speed of light, $\omega_B = eB/(m_e c)$ is the cyclotron frequency with $B$ denoting the strength of the local magnetic field, and $\theta$ is the angle between the photon trajectory and the direction of the magnetic field line.

The effectiveness of the scattering process is quantified by the resonant scattering optical depth, $\tau_{\rm res}$, which can be estimated by the integral:
\begin{align} \label{eq:tau_res}
    \tau_{\rm res} &= \int n_e \sigma_{\rm res}(\omega) {\rm d}l \\ \nonumber
                   &\sim \int_{R_{\rm NS}}^r n_e \sigma_{\rm res}(\omega) {\rm d}r' \\ \nonumber
                   &= \frac{\pi^2 e^2 n_e}{m_e c} (1 + \cos^2{\theta}) \int_{R_{\rm NS}}^r \delta(\omega - \omega_B(r')) {\rm d}r' \\ \nonumber
                   &= \frac{\pi^2 e^2 n_e}{m_e c} (1 + \cos^2{\theta}) \frac{r}{3 \omega_B} \\ \nonumber
                   &= \tau_0 (1 + \cos^2{\theta}),
\end{align}
where the integral is performed along the line of sight, $l$, for simplicity taken to be the radial distance, $r$, from the stellar surface, $n_e$ is the density of the magnetospheric plasma (assumed to be constant), and $\tau_0$ is defined as:
\begin{align} 
    \tau_0 \equiv \frac{\pi^2 e^2 n_e r}{3 m_e c \omega_B}.
\end{align}
Note that $\omega_{B}$ depends on the distance $r$, since in the simplified case of a poloidal dipole field, we have that $B(r) \sim B_{\rm P} R_{\rm NS}^3 / r^3$, where $B_{\rm P}$ is the magnetic field strength at the pole. To solve the integral in Eq.~\eqref{eq:tau_res}, we took advantage of the following property of the Dirac delta function:
\begin{align} 
    \int_{-\infty}^{+\infty} \delta(f(x)) {\rm d}x = \sum_i \frac{1}{\left| f'(x_i) \right|},
\end{align}
where $x_i$ are the zeros of the function $f(x)$. In our case, we have a function $f(r) \equiv \omega - \omega_B(r)$ and its derivative with respect to $r$ is $3 \omega_B / r$. The zeros of $f(r)$ are thus the values of the radial distance, $r$, where $\omega = \omega_B(r)$, i.e., where the frequency of the photon matches the cyclotron frequency.

To compute the spectrum distorted by resonant cyclotron scattering, we use the simplified semi-analytical 1D model described in \cite{Lyutikov2006} \citep[see also][]{Rea2008}. In this model, we assume that the seed thermal photons from the stellar surface propagate in the radial direction. They interact with the magnetospheric electron-positron plasma where particles are gyrating along the magnetic field lines and are assumed to have a top-hat thermal velocity distribution centered at zero and extending up to velocities $\pm \beta_{\rm T}$. In the 1D picture, we also have that photons propagate parallel to the magnetic field lines, either away or towards the star. Hence, in Eq.~\eqref{eq:tau_res}, $\cos \theta = \pm 1$ and $\tau_{\rm res} = 2 \tau_0$.
Moreover, a photon with a given energy, $E_0 = \hbar \omega_0$, resonantly interacts with electrons that are gyrating with a cyclotron frequency that matches the photon's energy. This defines the distance from the star, $r$, at which the interaction will happen.
The probabilities, $p_{+}$ and $p_{-}$, that a photon with $E_0$ undergoes resonant scattering and is transmitted or reflected, respectively, are given by Eq. (35) in \citet{Lyutikov2006}, which we report here:
\begin{align} 
    p_{+}(\eta){\rm d}\eta &= e^{-\tau_0/2} \left[ \delta(\eta) + \frac{\tau_0}{8 \beta_{\rm T}} \sqrt{\frac{4 \beta_{\rm T} - \eta}{\eta}} I_1 \left( \frac{\tau_0}{4 \beta_{\rm T}} \sqrt{\eta (4 \beta_{\rm T} - \eta)} \right)  \right]{\rm d}\eta, \\ \nonumber
    p_{-}(\xi){\rm d}\xi &= \frac{\tau_0}{8 \beta_{\rm T}} e^{-\tau_0/2} I_0 \left( \frac{\tau_0}{4 \beta_{\rm T}} \sqrt{(2 \beta_{\rm T} - \xi) (\xi + 2\beta_{\rm T})} \right){\rm d}\xi,
\end{align}
where $I_0$ and $I_1$ are the modified Bessel functions of the first kind of orders 0 and 1, respectively, $\eta = (E - E_0)/E_0$ and $\xi = (E_0 - E)/E_0$ with $E$ representing the energy of the scattered photon. Rewriting the probabilities above in terms of the energies gives the expressions \citep[see Eq. (36) in][]{Lyutikov2006}:
\begin{align} 
    p_{+}(E, E_0){\rm d}E_0 &= e^{-\tau_0/2} \left[ \delta(E - E_0) + \frac{\tau_0}{8 \beta_{\rm T} E_0} \sqrt{\frac{4 \beta_{\rm T} E_0 - E + E_0}{E - E_0}} \right. \\ \nonumber
    &\times \left. I_1 \left( \frac{\tau_0}{4 \beta_{\rm T} E_0} \sqrt{(E - E_0) (4 \beta_{\rm T} E_0 - E + E_0)} \right)  \right] {\rm d}E_0,\\[1.3ex]
    p_{-}(E, E_0){\rm d}E_0 &= \frac{\tau_0}{8 \beta_{\rm T} E_0} e^{-\tau_0/2}\\\nonumber
    &\times I_0 \left( \frac{\tau_0}{4 \beta_{\rm T} E_0} \sqrt{(2 \beta_{\rm T} E_0 - E_0 + E) (E_0 - E + 2\beta_{\rm T} E_0)} \right){\rm d}E_0.
\end{align}

Fig. 2 in \cite{Lyutikov2006} shows that transmitted photons can only gain energy, whereas reflected photons can both gain or lose energy. As the dipolar magnetic field decays as $\sim r^{-3}$, the cyclotron frequency of electrons gyrating along the magnetic field lines also decreases as $\sim r^{-3}$ as one moves away from the stellar surface. Therefore, a photon that is resonantly scattered and reflected back towards the stellar surface can only interact with the electron plasma again if it has gained energy during the scattering process. If this happens, the photon can be transmitted towards the observer. Because photons only gain energy during the transmission process, as they propagate away from the star, they do not interact with the electrons again because these have energies lower than those of the photons. 

The above probabilities depend on two free parameters, i.e., $\tau_0$ and $\beta_{\rm T}$. Following \citet{Gullon2015}, we assume that these two parameters depend on the dipolar magnetic field strength, $B$, according to the following scaling relations:
\begin{align} 
    \tau_0 &= \begin{cases} 0.001  \quad {\rm if} \, \unit[B \leq 10^{13}]{G}, \\ \frac{B}{\unit[10^{14}]{G}} \quad {\rm if} \, \unit[B > 10^{13}]{G}, \end{cases}
\end{align}
\begin{align} 
    \beta_{\rm T} = \begin{cases} 0.001  \quad {\rm if} \, \unit[B \leq 10^{13}]{G}, \\ 0.3 \quad {\rm if} \, \unit[B > 10^{13}]{G}. \end{cases}
\end{align}
These approximated relations roughly reproduce the correlations between $\tau_0$, $\beta_{\rm T}$, and $B$ found in \citet{Rea2008}.

If $n_{\rm BB}(E_0) = I_{\rm BB}(E_0)/E_0$ represents the thermal spectrum from the neutron star surface in photon counts, where
\begin{align} 
    I_{\rm BB}(E) = \frac{2}{h^3 c^2} \frac{E^3}{e^\frac{E}{k_{\rm B} \bar{T}_{\infty}} - 1}
\end{align}
denotes a black-body spectrum, the resulting photon intensity spectrum, $n_{\rm RCS}(E)$, due to cyclotron resonant scattering can be computed by summing the contributions coming from various reflection and transmission events, namely:
\begin{align} 
    n_{\rm RCS}(E) &= \int {\rm d} E_0 p_{+}(E, E_0) n_{\rm BB}(E_0) + \\ \nonumber
    &+ \int {\rm d} E_1 p_{+}(E, E_1) \int {\rm d} E_0 p_{-}(E_1, E_0) n_{\rm BB}(E_0) + \\ \nonumber
    &+ \int {\rm d} E_3 p_{+}(E, E_3) \int {\rm d} E_2 p_{-}(E_3, E_2)\int {\rm d} E_1 p_{-}(E_2, E_1) \\ \nonumber
    & ... \, .
\end{align}
This sum quickly converges and for numerical purposes, we follow \cite{Lyutikov2006} and consider six reflection terms only.

\section{TSNPE results on the full population}
\label{app:posterior_full_pop}

\begin{figure*}
\centering
\includegraphics[width = 1\textwidth]{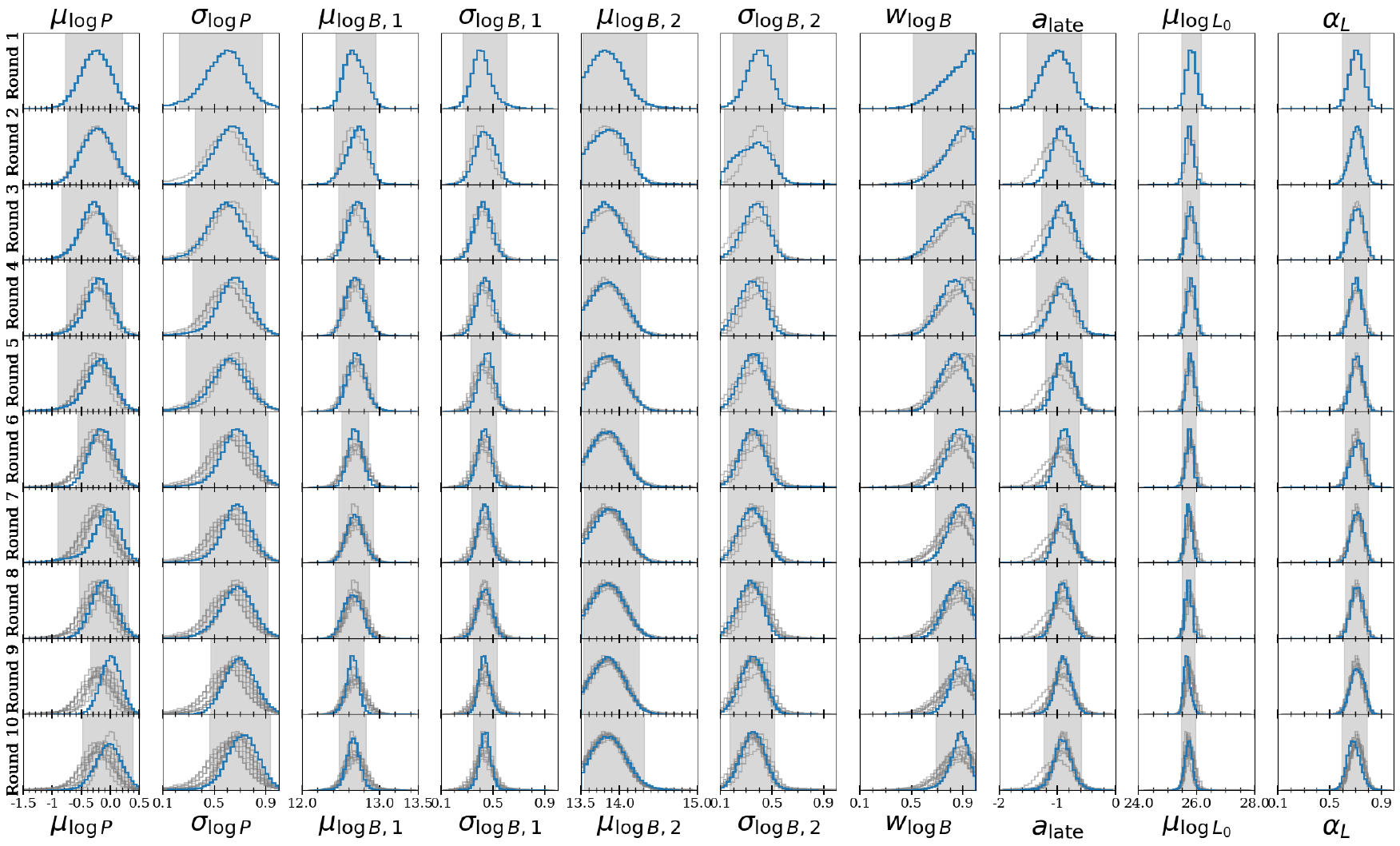}
\caption{Results of the \ac{TSNPE} algorithm applied to our population synthesis considering the entire X-ray sample (see Section~\ref{sec:results_allpop}). Each row shows the marginalized posteriors obtained in each round of inference while the columns refer to different parameters describing the log-normal birth spin-period distribution ($\mu_{\log P}$, $\sigma_{\log P}$), the double log-normal birth magnetic-field distribution ($\mu_{\log B,1}$, $\sigma_{\log B,1}$, $\mu_{\log B,2}$, $\sigma_{\log B,2}$, $w_{\log B}$), the power-law index for the late time decay of the magnetic field ($a_{\rm late}$), and the radio luminosity prescription ($\mu_{\log L_0}$, $\alpha_L$). For a given parameter, the marginalized posterior computed in that specific round is shown in blue, while the marginalized posteriors from previous rounds are shown in light gray. The gray shaded area represents the $95\%$ credibility interval of the approximated posterior for a given round. In each of these 1D marginal posterior distribution, the horizontal axes represent the parameters’ prior ranges.}
\label{fig:marginal_posterior_B_double_lognorm_dip-tor_heavy_full}
\end{figure*}

\begin{figure*}
\centering
\includegraphics[width = \textwidth]{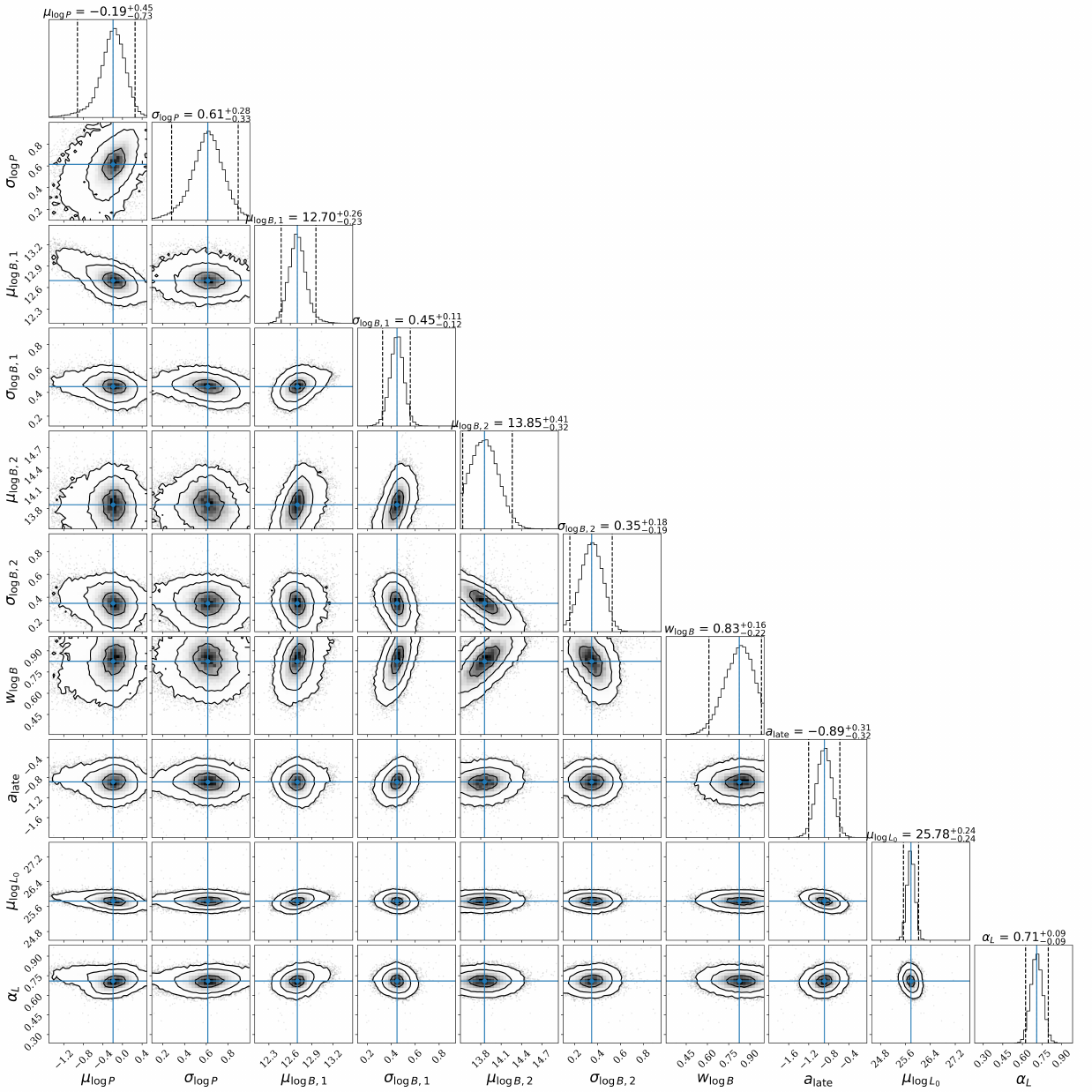}
\caption{Corner plot of the posterior distribution conditioned on the observed data including the radio pulsars and the entire X-ray sample (see Section~\ref{sec:results_allpop}). The posterior is estimated from round 5 of the TSNPE algorithm and adopted for the best-parameter analysis in Section~\ref{sec:results_allpop}. The parameters describe the log-normal birth spin-period distribution ($\mu_{\log P}$, $\sigma_{\log P}$), the double log-normal birth magnetic-field distribution ($\mu_{\log B,1}$, $\sigma_{\log B,1}$, $\mu_{\log B,2}$, $\sigma_{\log B,2}$, $w_{\log B}$), the power-law index for the late time decay of the magnetic field ($a_{\rm late}$), and the radio luminosity prescription ($\mu_{\log L_0}$, $\alpha_L$). The contour levels in the 2D marginalized posteriors represent the 1-$\sigma$, 2-$\sigma$ and 3-$\sigma$ credible regions. For each parameter, we report the median value and the 2-$\sigma$ credible interval. In the 1D marginalized posterior distributions, these are marked by the blue solid lines and the dashed lines, respectively.}
\label{fig:corner_plot_full}
\end{figure*}

Here, we report the results of the TSNPE algorithm for the experiment where we train on the full samples of radio pulsars and X-ray emitting neutron stars as described in Section~\ref{sec:results_allpop}. In Fig.~\ref{fig:marginal_posterior_B_double_lognorm_dip-tor_heavy_full}, we show the marginal posteriors obtained in the ten training rounds. In Fig.~\ref{fig:corner_plot_full}, we show the corner plot of the posterior obtained in round 5 used to extract the best parameters displayed in Table~\ref{tab:infer_results}. 


\section{Predictive check and coverage test}
\label{app:predictive_coverage_test}

\begin{figure*}
\centering
\includegraphics[width = \textwidth]{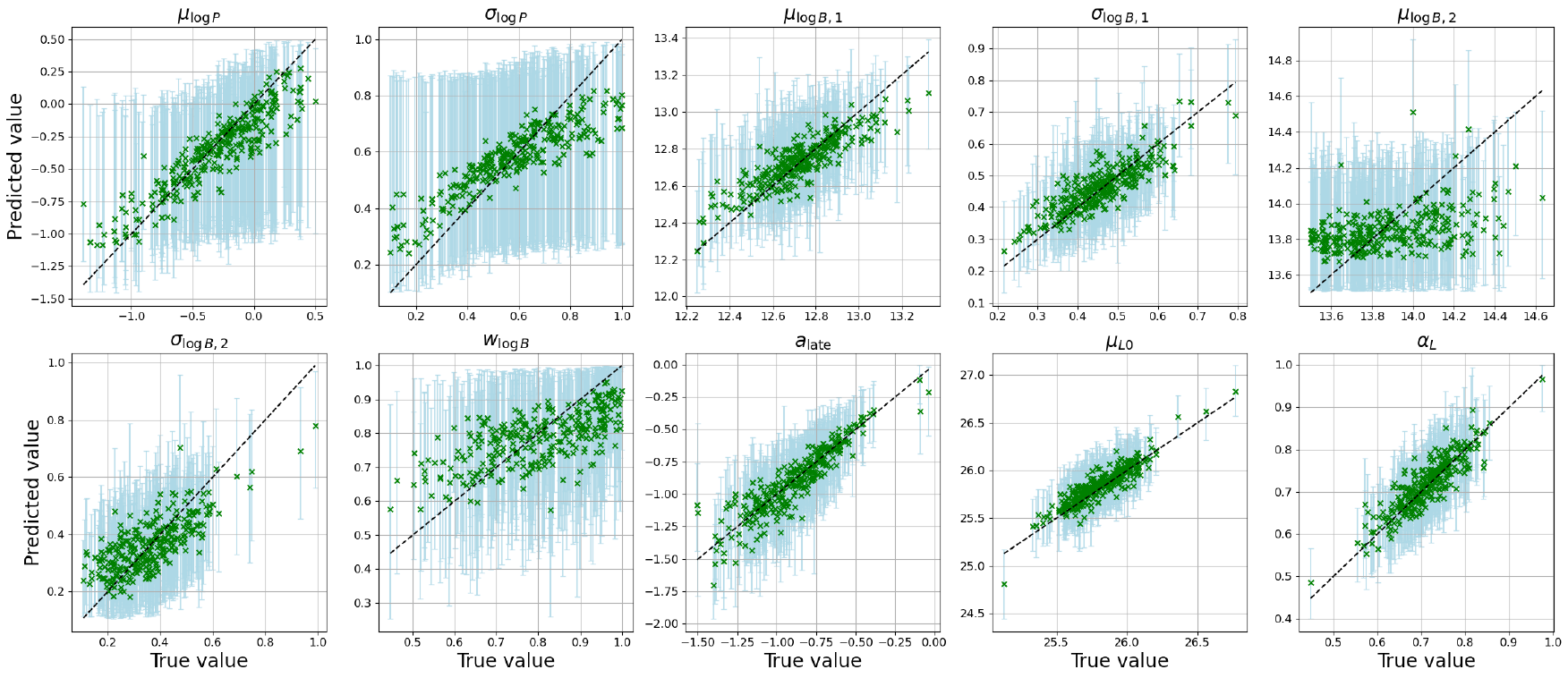}
\caption{Results of posterior predictive checks performed on a test dataset of 300 simulations using the density estimator trained on radio pulsars and the full X-ray neutron star sample (see Section~\ref{sec:results_allpop}). We use the density estimator obtained from round 5 (see also Fig.~\ref{fig:marginal_posterior_B_double_lognorm_dip-tor_heavy_full}). For each parameter, we display the predicted values and their uncertainties, i.e., the median values (green crosses) and the 95\% credible interval (light blue error bars) of the posterior distribution on the y-axis and the true value on the x-axis. The black dashed lines indicate where the predicted value is equal to the true value.}
\label{fig:predictive_test_full}
\end{figure*}

\begin{figure*}
\centering
\includegraphics[width = \textwidth]{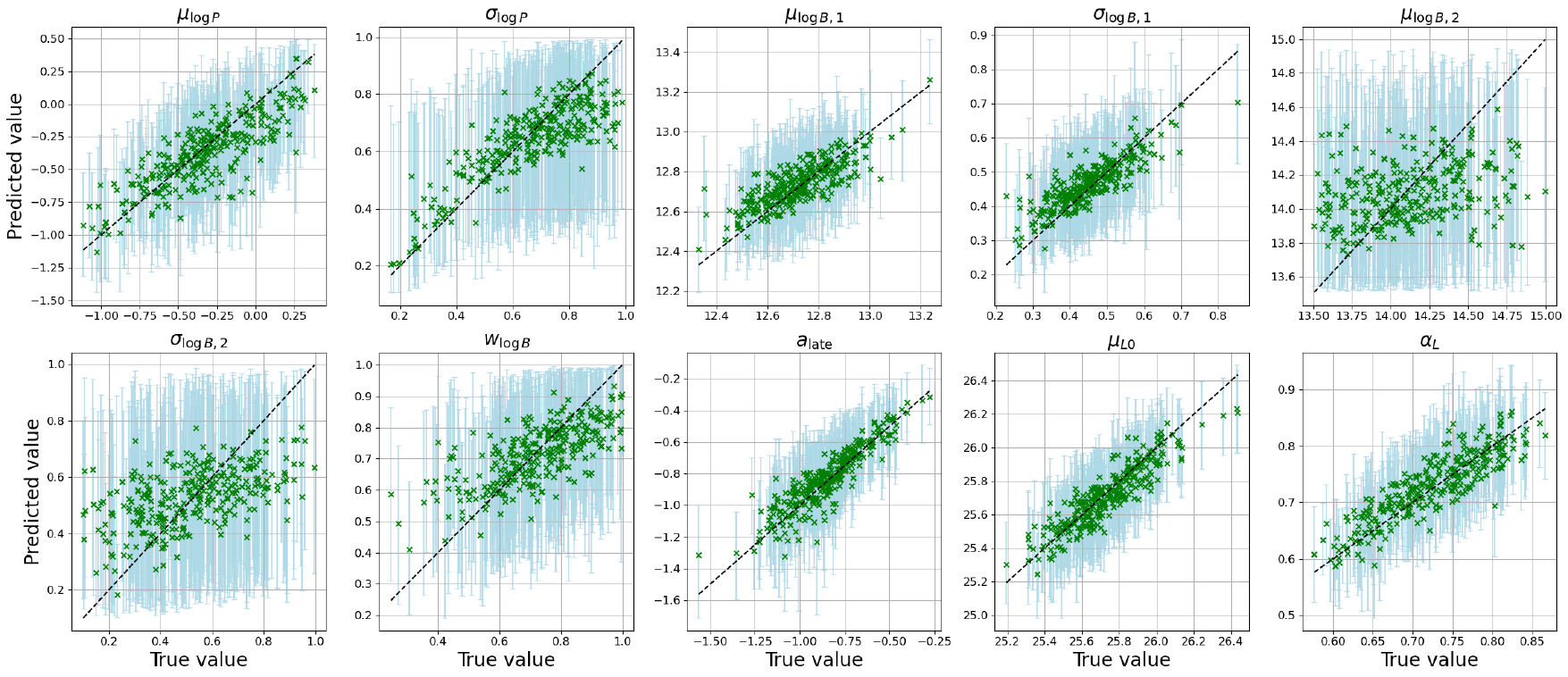}
\caption{Results of posterior predictive checks performed on a test dataset of 300 simulations using the density estimator trained on radio pulsars and the sub-populations of young magnetars and XDINSs (see Section~\ref{sec:results_youngxdins}). We used the density estimator obtained from round 5 (see also Fig.~\ref{fig:marginal_posterior_B_double_lognorm_dip-tor_heavy_youngxdins}). For each parameter, we display the predicted values and their uncertainties, i.e., the median values (green crosses) and the 95\% credible interval (light blue error bars) of the posterior distribution on the y-axis and the true value on the x-axis. The black dashed lines indicate where the predicted value is equal to the true value.}
\label{fig:predictive_test_youngxdins}
\end{figure*}

\begin{figure*}
\centering
\includegraphics[width = 0.9\textwidth]{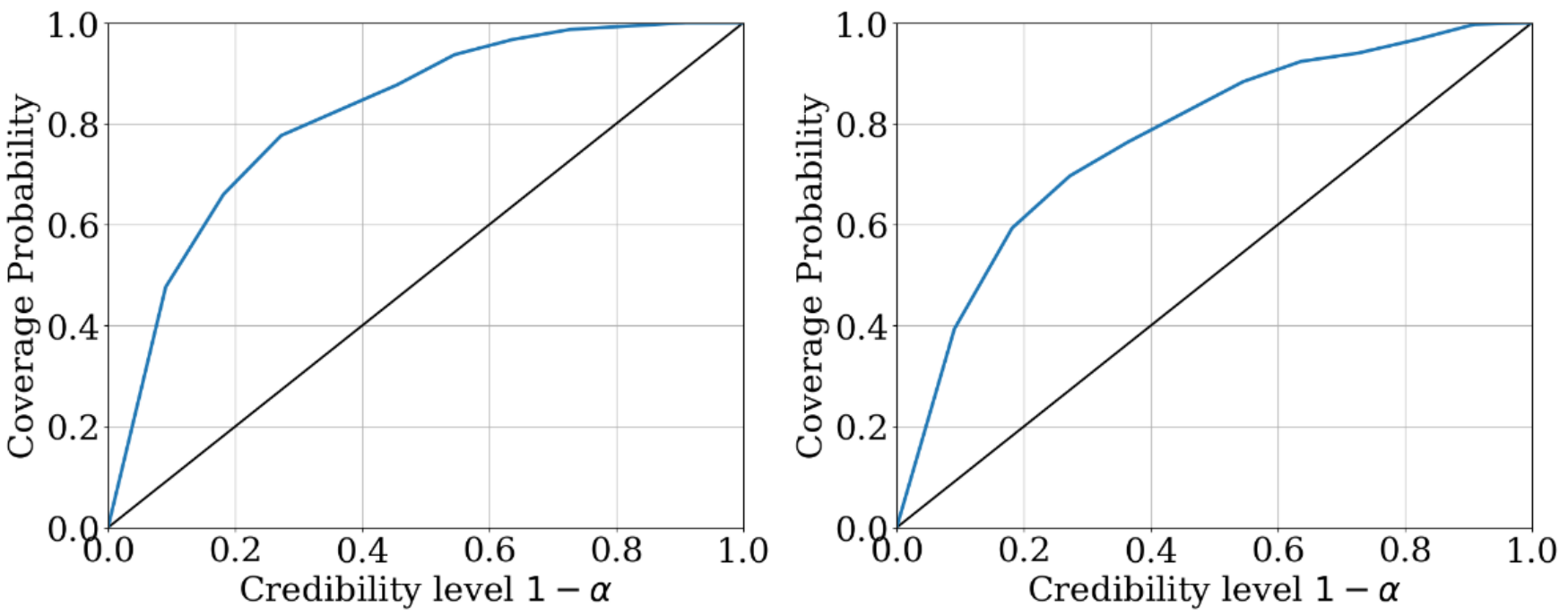}
\caption{Coverage diagnostic plots for the posterior estimators trained on radio pulsars and the full X-ray sample (left panel, see Section~\ref{sec:results_allpop}) and on radio pulsars and the sub-populations of young magnetars and XDINSs (right panel, see Section~\ref{sec:results_youngxdins}). In both cases, we use the posterior estimator obtained from round 5 of the TSNPE algorithm.}
\label{fig:coverage_plots}
\end{figure*}

To evaluate the reliability of the trained posterior estimators in round 5 for the experiments described in Sections~\ref{sec:results_allpop} and~\ref{sec:results_youngxdins}, we check the estimator performance on a test dataset of 300 simulations drawn from the proposal restricted prior. In Figs.~\ref{fig:predictive_test_full} and~\ref{fig:predictive_test_youngxdins}, we show the inference results on the test dataset for each of the ten parameters we are inferring for the two experiments, respectively. For each test sample, we report the median value and the 95\% credible interval from the predicted posterior distribution and compare it with the ground truth value.
Overall, we recover behavior that follows the diagonal indicating that the network shows good performance when recovering the ground-truth values. However, we observe that the network struggles the most when inferring $\mu_{\log P}$ and $\sigma_{\log P}$, which describe the initial spin-period distribution as well as $\mu_{\log B,2}$, $\sigma_{\log B,2}$, and $w_{\log B}$, which characterize the initial magnetic-field distribution (see Sections~\ref{sec:results_allpop} and~\ref{sec:discussion} for further interpretation). 
We also perform the coverage probability test \citep{Cook2006, Hermans2021} on the trained estimators (see Fig.~\ref{fig:coverage_plots}). To do this, we use the 300 test samples and compute the percentage of ground truth values that fall inside a given credibility level of the corresponding estimated posteriors. The coverage plots show that the posterior estimators are overall conservative, i.e., the coverage line lie well above the diagonal \citep[see also][for an explanation of the coverage probability]{Graber2024, Pardo-Araujo2025}.


\bibliography{bibliography}{}
\bibliographystyle{aasjournal}

\end{document}